\documentclass[twoside,11pt]{article}

\usepackage{jmlr2e}
\usepackage{amsmath}
\usepackage{subfigure}
\usepackage{lipsum,pdflscape}
\usepackage[utf8]{inputenc}
\usepackage{caption}
\usepackage{float}
\usepackage{makecell}
\usepackage{diagbox}
\usepackage{multirow}
\usepackage{algpseudocode}
\usepackage{algorithm}
\usepackage{verbatim}
\newenvironment{nospaceflalign*}
 {\setlength{\abovedisplayskip}{0pt}\setlength{\belowdisplayskip}{2.0pt}%
  \csname flalign*\endcsname}
 {\csname endflalign*\endcsname\ignorespacesafterend}

\firstpageno{1}

\begin{document}

\title{Neural Operators Learn the Local Physics of Magnetohydrodynamics}

\author{\name Taeyoung Kim \email legend@snu.ac.kr \\
       \addr Department of Mathematical Sciences\\
       Seoul National University\\
       Seoul 08826, South Korea
       \AND
       \name Youngsoo Ha \email youngamath@snu.ac.kr \\
       \addr Department of Mathematical Sciences\\
       Seoul National University\\
       Seoul 08826, South Korea
       \AND
       \name Myungjoo Kang \email mkang@snu.ac.kr \\
       \addr Department of Mathematical Sciences\\
       Seoul National University\\
       Seoul 08826, South Korea}
       
\maketitle

\begin{abstract}%   <- trailing '%' for backward compatibility of .sty file
Magnetohydrodynamics (MHD) plays a pivotal role in describing the dynamics of plasma and conductive fluids, essential for understanding phenomena such as the structure and evolution of stars and galaxies, and in nuclear fusion for plasma motion through ideal MHD equations. Solving these hyperbolic PDEs requires sophisticated numerical methods, presenting computational challenges due to complex structures and high costs. Recent advances introduce neural operators like the Fourier Neural Operator (FNO) as surrogate models for traditional numerical analyses. This study explores a modified Flux Neural Operator (Flux NO) model to approximate the numerical flux of ideal MHD, offering a novel approach that outperforms existing neural operator models by enabling continuous inference, generalization outside sampled distributions, and faster computation compared to classical numerical schemes.
\end{abstract}


\section{Introduction}

\subsection{Magnetohydrodynamics}
In the mid-20th century, the advent of the Space Age and growing interest in nuclear fusion led to an intensified focus and understanding of plasma. During this period, significant theoretical studies on plasma were conducted, marking early research in kinetic equations by (\cite{Vlasov:38}) and studies on magnetohydrodynamics (MHD) by (\cite{Alfven:42}). MHD describes the behavior of electrically conductive fluids and is primarily used for plasmas and liquid metals under the influence of large-scale, low-frequency magnetic fields (\cite{Bittencourt:04}). The applications of MHD extend across various fields, including astrophysics (\cite{Kennel:85}), solar physics (\cite{Priest:82}), the study of Earth's magnetosphere (\cite{Mukhopadhyay:21}), and research in nuclear fusion (\cite{Wesson:78}). Ideal MHD represents one of the simplest forms of these models, disregarding dissipative effects such as viscosity, thermal conductivity, and resistance. Despite its simplicity, ideal MHD proves to be a powerful tool capable of explaining a wide range of plasma phenomena (\cite{Galtier:16}). This includes its use in analyzing plasma within tokamaks and stellarators, understanding the dynamo process generating Earth's magnetic field (\cite{Rincon:19}), studying plasma phenomena inside and on the surface of the sun (\cite{Shibata:11}), and describing the formation of galactic structures and the universe (\cite{Pakmor:13}).
\noindent

\subsection{Numerical Schemes}

MHD can be viewed as a generalized model that combines hydrodynamics with electromagnetism. In particular, ideal MHD is considered as a hyperbolic conservation law, conserving mass, momentum, magnetic fields, and energy density. Numerical methods for solving hyperbolic conservation laws have a long history, with various approaches proposed to achieve stability, computational efficiency, and high-order accuracy. High-order accurate methodologies, such as essentially non-oscillatory (ENO) (\cite{Harten:87}) and weighted ENO (WENO) (\cite{Liu:94}) schemes, have been effective in solving hyperbolic conservation laws. Jiang and Shu introduced the WENO scheme with third and fifth-order accuracy (WENO-JS) (\cite{Jiang:96}), noted for its robust shock-capturing capability, albeit with a disadvantage of being dissipative in turbulent flows. To address this shortcoming, Henrick et al. developed the enhanced WENO-M scheme (\cite{Henrick:05}). However, applying WENO schemes to the system of conservation laws involves significant computational costs. Various attempts to reduce computational costs include Jiang and Shu's work (\cite{Jiang:96}), Pirozzoli's hybrid compact-WENO scheme (\cite{Pirozzoli:02}), and the efforts by (\cite{Hill:04})(\cite{Costa:07}) among others. For ideal MHD, maintaining the divergence-free condition for the magnetic field is essential, with methods like the Constrained Transport (CT) method by (\cite{Evans:88}), and the projection scheme by (\cite{Brackbill:80}) proposed. Comprehensive reviews on methods ensuring the divergence-free condition in ideal MHD are available in (\cite{Toth:00}). Early studies on numerical solutions for ideal MHD include works by (\cite{Brio:88}), (\cite{Dai:98}), (\cite{Jiang:99}). And other various efforts to solve ideal MHD numerically have been develeoped (\cite{Christlieb:14})(\cite{Rossmanith:06}). In this research, we innovatively combine traditional numerical schemes with artificial neural network techniques to reduce computational costs and enforce the divergence-free condition for ideal MHD problems.

\subsection{Neural Operators and Flux Neural Operator}
In recent times, several Neural Operators have been proposed as a method to replace conventional numerical analysis techniques with machine learning approaches. Notably, the Graph Kernel Network (\cite{Li:20}), Fourier Neural Operator (FNO) (\cite{Li:21}), DeepONet (\cite{Lu:21}), and various adaptations thereof have been introduced (\cite{Gupta:21})(\cite{Wen:22})(\cite{Lee:23}). A common feature among these models is their ability to handle functional data, as they are not limited by the resolution of input data. This characteristic allows for the approximation of operators—mappings between functional spaces—and paves the way for these models to serve as surrogate models for numerical schemes. Unlike traditional numerical schemes that iteratively calculate solutions, most neural operator models learn global solvers that map initial data directly to the solution data at a specific time. One advantage of this method is its significantly faster computation speed compared to classical numerical schemes (\cite{Pathak:22}); however, it often lacks generalization capability (\cite{Kim:24}), making it challenging to learn actual broad physical phenomena. The recently proposed Flux FNO model (\cite{Kim:24}) combines the strengths of both numerical schemes and neural operators. In the respective study, it demonstrated a leap in generalization capability by learning numerical flux for hyperbolic conservation law problems, which enabled the neural operator model to learn local physics. While the study empirically showed the approximation of numerical flux by Flux FNO for one-dimensional scalar conservation laws, our paper takes this a step further by applying it to one of the most challenging problems, the ideal MHD problem.

\subsection{Our contribution}
In this paper, we employ various techniques motivated by the physical properties of equations and numerical analysis to enhance the recently developed machine learning method, Flux FNO, for solving the ideal MHD problem. As the problem transitions from scalar-valued to vector-valued outputs, we have redesigned the Flux FNO model architecture. This redesign allows the model to process each physical variable—density, velocity, magnetic field, energy—separately, addressing the experimental findings that revealed limitations in the model’s expressiveness when handling all variables simultaneously. Secondly, we design and apply a loss function that confers the Total Variation Diminishing (TVD) property to the approximated numerical flux, ensuring stability. Unlike classical numerical schemes, merely approximating the numerical flux without such measures could lead to severe oscillations as iterations accumulate. Furthermore, we enforce the divergence-free condition by implementing a loss specific to the divergence of magnetic field variables. This specialized loss imparts an inductive bias suitable for the ideal MHD solver, effectiveness of which we confirm through an ablation study. Utilizing these methodologies, our experiments are designed to qualitatively assess its generalization performance, including inference over continuous time and on out-of-distribution samples. We also apply it to solve representative test problems of ideal MHD, comparing its effectiveness with traditional numerical methodologies.

\section{Preliminaries}

\subsection{Ideal MHD Equations}
Ideal MHD is described by a system of coupled partial differential equations that characterize conducting fluids. Among various formulations of MHD, the ideal MHD equations represent the simplest form, embodying a synthesis of fluid dynamics and Maxwell's equations, while excluding effects such as viscosity, resistance, and thermal conductivity. The conservative form of ideal MHD can be articulated as follows:

\begin{align*}
    \rho_{t}+\nabla \cdot(\rho\bold{u})&=0,\\
    (\rho\bold{u})_{t}+\nabla \cdot\Bigg[\rho\bold{u}\otimes\bold{u}+\Bigg(p+\frac{1}{2}\|\bold{B}\|^{2}\Bigg)\bold{I}-\bold{B}\otimes\bold{B}\Bigg]&=0, \\
    \bold{B}_{t}+\nabla\cdot(\bold{u}\otimes\bold{B}-\bold{B}\otimes\bold{u})&=0, \\
    \mathcal{E}_{t}+\nabla\cdot\Bigg[\Bigg(\mathcal{E}+p+\frac{1}{2}\|B\|^{2}\Bigg)\bold{u}-\bold{B}(\bold{u}\cdot\bold{B})\Bigg]&=0.
\end{align*}
\noindent
where $\bold{u}$ represents velocity, $\bold{B}$ the magnetic field, $\rho$ density, $E$ energy, and $p$ pressure. Additionally, according to Maxwell’s equations, the magnetic field must satisfy the divergence-free condition, which is expressed as:

\begin{align*}
    \nabla\cdot\bold{B}=0.
\end{align*}
\noindent
In the scenario where the fluid is considered incompressible, an analogous divergence-free condition for velocity would be necessary. However, throughout this paper, we focus on compressible fluids, where such a condition for velocity is not explicitly required due to the fluid's capacity to vary in density. The pressure and energy is coupled through the following equation:
\begin{align*}
    p=(\gamma-1)\bigg(\mathcal{E}-\frac{1}{2}\rho\bold{u}^{2}-\frac{1}{2}\|\bold{B}\|^{2}\Bigg).
\end{align*}
\noindent
where $\gamma$ is the ratio of specific heats.

\noindent
{\bf One dimensional ideal MHD}{
The governing equation for ideal MHD in the one-dimensional case, expressed in conservative form, can be written as follows:

\begin{align*}
    \frac{\partial}{\partial t}\begin{bmatrix}
        \rho\\
        \rho u_{x}\\
        \rho u_{y}\\
        \rho u_{z}\\
        B_{y}\\
        B_{z}\\
        \mathcal{E}
    \end{bmatrix} +
    \frac{\partial}{\partial x}\begin{bmatrix}
        \rho u_{x} \\
        \rho u_{x}^{2} + p^{*} - B_{x}^{2} \\
        \rho u_{x}u_{y}-B_{x}B_{y}\\
        \rho u_{x}u_{z}-B_{x}B_{z}\\
        B_{y}u_{x}-B_{x}u_{y}\\
        B_{z}u_{x}-B_{x}u_{z}\\
        (\mathcal{E}+p^{*})u_{x}-B_{x}(\bold{u}\cdot\bold{B})
    \end{bmatrix}
    =0.
\end{align*}
\noindent
where $p^{*}=p+\frac{\bold{B}^{2}}{2}$ represents the total pressure, incorporating magnetic pressure. Each variable with a subscript denotes components of velocity and magnetic field.

}
\noindent
{\bf Two dimensional ideal MHD}{
For two-dimensional case, the equtions are expressed in conservative form, can be written as follows:

\begin{align*}
    \frac{\partial}{\partial t}\begin{bmatrix}
        \rho\\
        \rho u_{x}\\
        \rho u_{y}\\
        \rho u_{z}\\
        B_{x}\\
        B_{y}\\
        B_{z}\\
        \mathcal{E}
    \end{bmatrix} +
    \frac{\partial}{\partial x}\begin{bmatrix}
        \rho u_{x} \\
        \rho u_{x}^{2} + p^{*} - B_{x}^{2} \\
        \rho u_{x}v_{y}-B_{x}B_{y}\\
        \rho u_{x}v_{z}-B_{x}B_{z}\\
        0\\
        B_{y}u_{x}-B_{x}u_{y}\\
        B_{z}u_{x}-B_{x}u_{z}\\
        (\mathcal{E}+p^{*})u_{x}-B_{x}(\bold{u}\cdot\bold{B})
    \end{bmatrix} +
    \frac{\partial}{\partial y}\begin{bmatrix}
        \rho u_{y}\\
        \rho u_{y}u_{x}-B_{y}B_{x}\\
        \rho u_{y}^{2} + p^{*} - B_{y}^{2}\\
        \rho u_{y}u_{z}-B_{y}B_{z}\\
        B_{x}u_{y}-B_{y}u_{x}\\
        0\\
        B_{z}u_{y}-B_{y}u_{z}\\
        (\mathcal{E}+p^{*})u_{y}-B_{y}(\bold{u}\cdot\bold{B})
    \end{bmatrix}
    =0.
\end{align*}

}
\subsection{Numerical schemes}{
To apply our methodology, a dataset consisting of discretized functions over time is required. Although it is possible to generate the dataset using actual observations, we created both training and testing datasets using classical numerical analysis techniques. This section describes the numerical methods utilized to generate the datasets, along with concepts related to the stability of numerical solutions.

\noindent
{\bf The WENO schemes}{
When using high-order numerical schemes, the phenomenon of oscillations at discontinuities, known as Gibbs phenomena, occurs. To address this issue, methods such as flux limiters, essentially non-oscillatory schemes (ENO), and slope limiters have been devised. Among these methods, the Weighted Essentially Non-Oscillatory (WENO) scheme stands out as it reconstructs the function values in a non-oscillatory manner by nonlinearly weighting each sub-stencil, utilizing the given function values. As an example, let's apply this to the one-dimensional conservation laws:

\begin{align}
    u_{t}+f(u)_{x}=0,\quad x\in\mathbb{R},\quad t\geq0. \tag{1} \label{eq:1}
\end{align}
\noindent
Let $x_{0}<\dots<x_{n}$ be the uniform discretization of the computational domain. where $ x_{j+\frac{1}{2}}=\frac{x_{j}+x_{j+1}}{2}$. The equation \eqref{eq:1} can be approximated with semi-discrete conservation schemes:

\begin{align}
    \frac{du_{j}}{dt}=-\frac{\partial f}{\partial x}\Big|_{x=x_{j}}.  \tag{2} \label{eq:2}
\end{align}
\noindent
where $u_{j}(t)$ is numerical approximation of function $u(x_{j},t)$ on a grid. And by approximating right term of \eqref{eq:2} in a conservative manner, we get following formula:

\begin{align}
    \frac{du_{j}}{dt}=-\frac{\hat{f}_{j+\frac{1}{2}}-\hat{f}_{j-\frac{1}{2}}}{\Delta x}.  \tag{3} \label{eq:3}
\end{align}
\noindent
where $\hat{f}_{j\pm\frac{1}{2}}$ is a numerical flux which satisfies lipschitz continuity and consistency with the physical flux $f$, namely, $\hat{f}(u,\dots,u)=f(u)$.
In fifth-order finite differnece WENO scheme (WENO-JS), the numerical flux is constructed using 5-point stencil which is subdivided into three sub-stencils. The computation of WENO-JS is as follows:

\begin{align*}
    \hat{f}_{j+\frac{1}{2}}=\sum_{k=0}^{2}\omega_{k}\hat{f}_{k,j+\frac{1}{2}}.
\end{align*}
\noindent
which is weighted sum of numerical fluxes $\hat{f}_{k,j+\frac{1}{2}}(k=0,1,2)$ which are as follows:

\begin{align*}
    \hat{f}_{0,j+\frac{1}{2}}=\frac{1}{3}f_{j-2}-\frac{7}{6}f_{j-1}+\frac{11}{6}f_{j},\\
    \hat{f}_{1,j+\frac{1}{2}}=-\frac{1}{6}f_{j-1}+\frac{5}{6}f_{j}+\frac{1}{3}f_{j+1},\\
    \hat{f}_{2,j+\frac{1}{2}}=\frac{1}{3}f_{j}-\frac5{6}f_{j+1}-\frac{1}{6}f_{j+2}.
\end{align*}
\noindent
The nonlinear weights $\omega_{k}$ are contingent upon the variant WENO construction techniques, with WENO-JS and WENO-Z being predominantly utilized. The detailed calculation of these weights can be found in (\cite{Jiang:96}) and (\cite{Borges:08}).
}

\noindent
{\bf Total Variation Diminish Runge-Kutta method}{
The total variation of numerical solution is defined as follows:

\begin{align*}
    TV(u)=\sum_{j}|u_{j+1}-u_{j}|.
\end{align*}
\noindent
And, we say numerical scheme has total variation diminishing (TVD) property if it satisfies following condition:

\begin{align*}
    TV(u^{n+1})\leq TV(u^{n}). \tag{4} \label{eq:4}
\end{align*}

}
\noindent
Let denote $-L(u)$ is approximation of spatial derivative $f(u)_{x}$ in \eqref{eq:2} then a general Runge-Kutta method for \eqref{eq:1} can be written as follows:

\begin{align}
    u^{0} &= u^{n}, \nonumber \\
    u^{i} &= \sum_{k=0}^{i-1}\Big(\alpha_{ik}u^{(k)}+\Delta t \beta_{ik}L(u^{(k)})\Big),\quad i=1,\dots,m, \tag{5} \label{eq:5}\\
    u^{n+1} &= u^{m}.  \nonumber
\end{align}
\noindent
It is known that Ruge-Kutta method \eqref{eq:5} is TVD under the following conditon (Courant-Friedrichs-Lewy (CFL) condition):

\begin{align*}
    \lambda&\leq\lambda_{0} \min_{i,k} \frac{\alpha_{ik}}{|\beta_{ik}|},\\
    \lambda &= \frac{\Delta t}{\Delta x}.
\end{align*}
\noindent
where $\lambda_{0}$ is a suitable CFL restriction.

}
\noindent
\subsection{Flux Neural Operator}
{\bf Fourier Neural Operator}{

\begin{figure}[htp!]
\centering
\includegraphics[height=7.5cm]{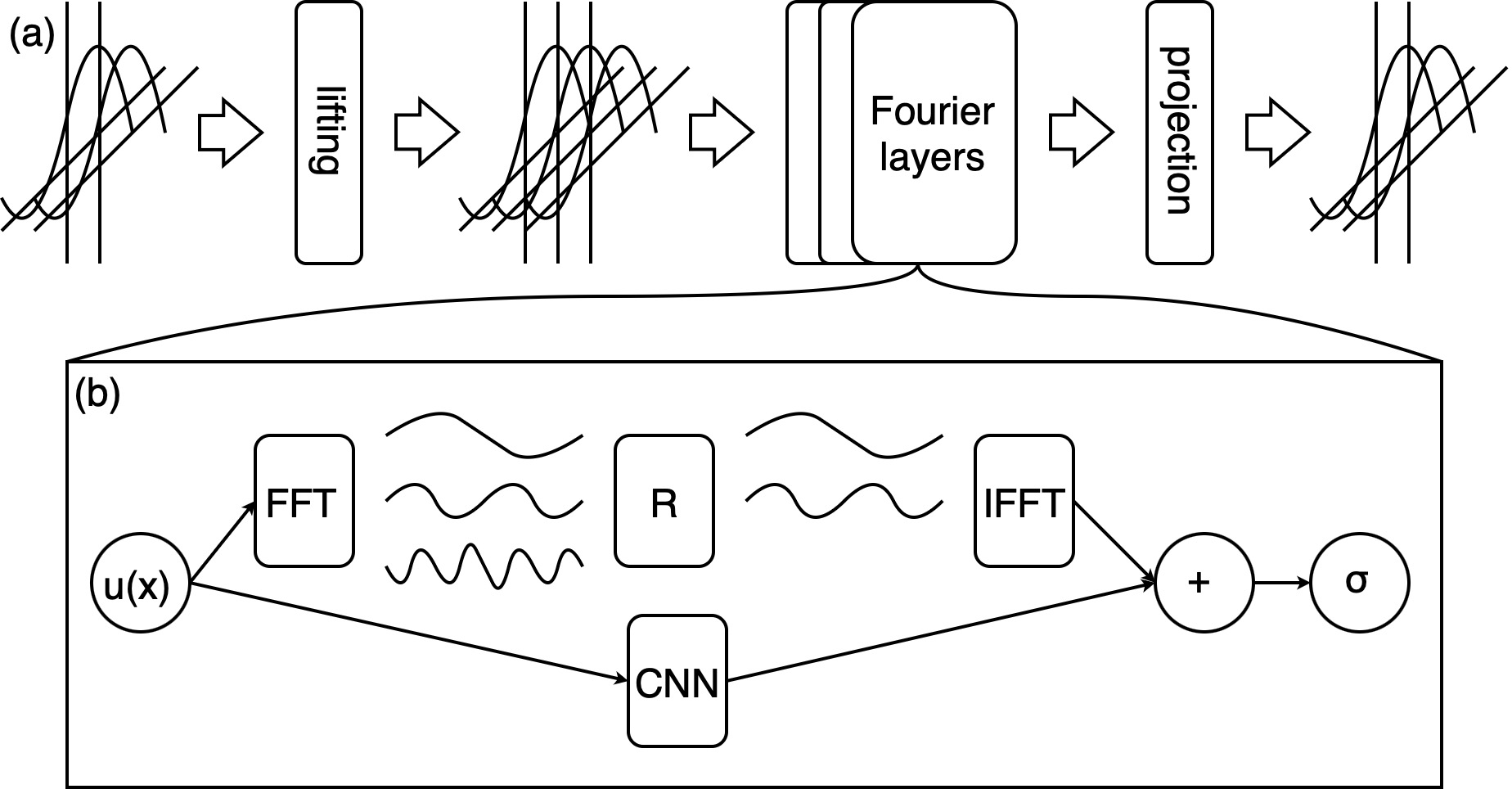}

\caption{Schematic of FNO with CNN layers}\label{FNO}
\end{figure}

\noindent
Among the various types of neural operators for handling functional data, the FNO is a neural operator model that processes the global convolution operations of functions by leveraging the Fourier transform. Its most notable feature is that the convolution with the kernel function is expressed through the Fourier transform as multiplication with a high-order tensor. The structure of the FNO is depicted in Figure \ref{FNO}, where the values of discretized functional data are non-linearly lifted by the lifting layer, passed through the Fourier layers, and finally projected into the desired dimension vector by the projection layer composed of FNCs. As shown in diagram (b) of Figure \ref{FNO}, each Fourier layer consists of a function convolution operation that globally transforms the data and an auxiliary neural network, which must maintain resolution invariance, with CNN layers commonly adopted to handle local data processing. The mathematical formulation for a FNO with a CNN layers can be represented as follows:

\noindent
{\bf Definition (FNO with CNN layers)} {
\begin{flalign*}
v_{0}&:=\mathcal{N}_{P}(a|_{X})=(\mathcal{N}_{P}(a_{\textbf{x}\cdot})_{j})_{\textbf{x}\in \textbf{X}, j=1,\dots,d_{v_{0}}}, \\
v_{t+1}&:=\mathcal{A}_{t+1}(v_{t})=\sigma\bigg(C_{t+1}(c_{1},\dots,c_{d})(\tilde{v_{t}})+\mathcal{F}^{-1}\Big(R_{t+1}\cdot(\mathcal{F}(v_{t}))\Big)\bigg)  \\
&=\sigma\Big(\sum_{k=1}^{d_{u}}\sum_{j_{1}=0}^{c_{1}-1}\cdots\sum_{j_{d}=0}^{c_{d}-1} K_{t+1,jk,j_{1},\dots,j_{d}}\tilde{v}_{t,x_{1}+j_{1},\dots,x_{d}+j_{d},k}
\\
&+ \sum_{\textbf{z},\textbf{k}\in K,k}  {D^{\dag}_{\textbf{x}\textbf{k}}R_{t+1,\textbf{k},jk}D_{\textbf{k}\textbf{z}}v_{t,\textbf{z}k}}\Big), \quad (t=0,...,L-1) \\
G(a;\theta)&:=\mathcal{N}_{Q}(v_{L})=(\mathcal{N}_{Q}(v_{L\textbf{x}\cdot})_{j})_{\textbf{x}\in \textbf{X}, j=1,\dots,d_{v_{L}}}.
\end{flalign*}
} 
Here, $\mathcal{N}_{P}$ and $\mathcal{N}_{Q}$ are neural networks used for lifting and projection, respectively, while $\mathcal{A}_{t+1}$ represents Fourier layers with $L$ being the depth of these Fourier layers. $a|_{X}$ represents the discretized functional data of $a$, $\sigma$ is an activation function, and each $C_{i}$ is a $d$-dimensional CNN layer with a kernel tensor $K_{i}$. The $R_{i}$ tensors are weight tensors that parameterize the kernel function of the Fourier layers. $D_{\textbf{k}\textbf{z}}$ represents the components of the discrete Fourier transform, also denoted as $\mathcal{F}$.

}

\noindent
{\bf Flux Neural Operator}{

\begin{figure}[htp!]
\centering
\includegraphics[height=7.5cm]{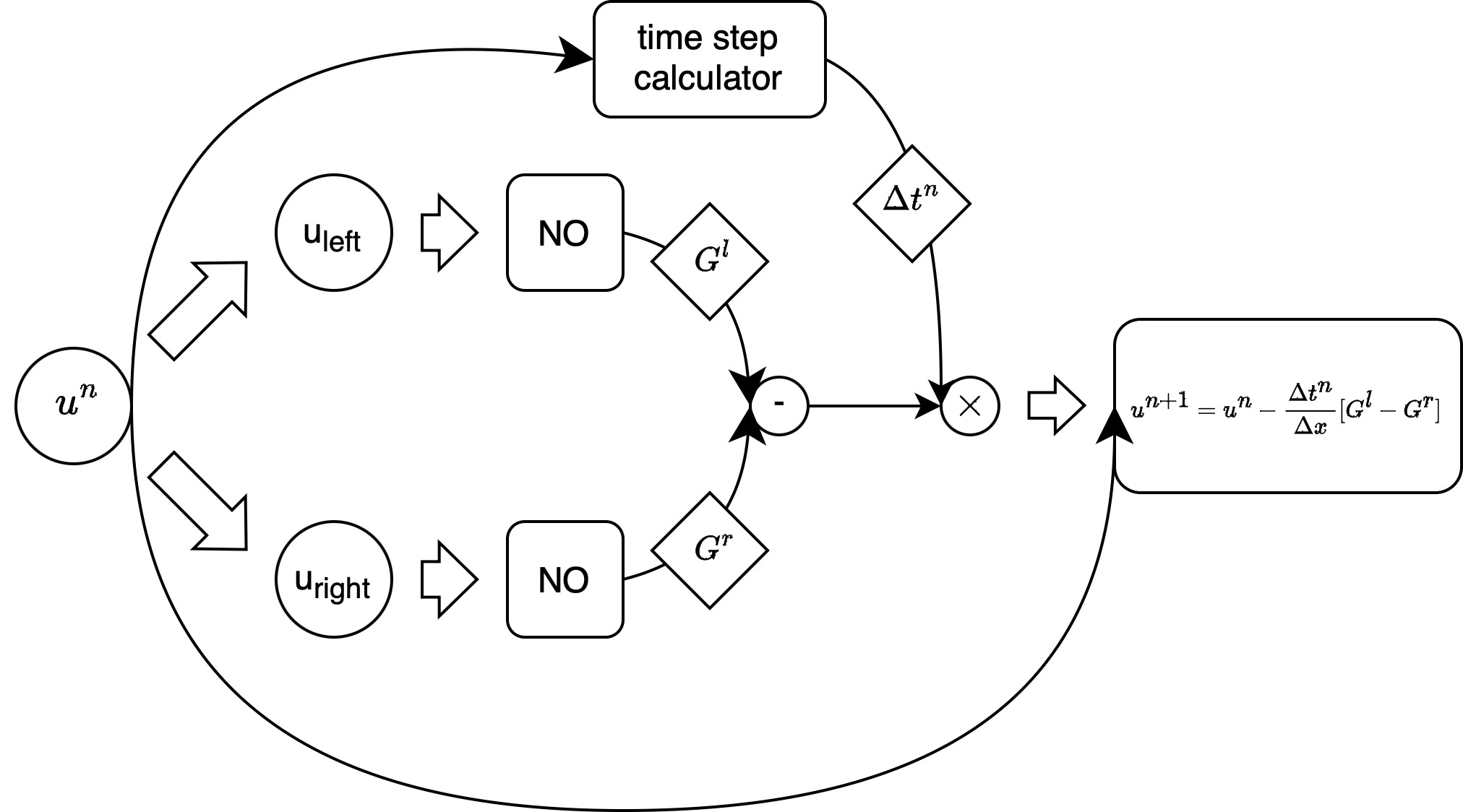}

\caption{Schematic of the forward structure for flux neural operator (Flux NO)}\label{FLUXNO}
\end{figure}

\noindent
The Flux Fourier Neural Operator (Flux FNO) introduced in the paper (\cite{Kim:24}), inspired by the numerical schemes of hyperbolic conservation laws, represents a method designed to learn the local physics of conservation laws directly through the approximation of fluxes, differing from original neural operators that predict target snapshots of solutions. The area where FNO is applied within Flux FNO can be substituted with other Neural Operators, leading to the generalized concept referred to as the Flux Neural Operator (Flux NO). The schematic of Flux NO for domain of one-dimensional case is shown in Figure \ref{FLUXNO}, and unlike conventional neural operators, Flux NO aims to approximate the flux itself and calculates local residuals based on this approximation, offering a novel approach to handling local physical phenomena. For an N-dimensional problem, the Flux NO operates as described in formula \eqref{eq:6}. Each $G_{i}(\cdot;\theta_{i})$ is approximated flux function along spatial direction $x_{i}$ ($\theta_{i}$ is parameter of neural operator). And each $\bold{U}_{j}^{l_{i,k}}$, $\bold{U}_{j}^{r_{i,k}}$ denote shifted $\bold{U}_{j}$ along spatial direction (which is corresponding spatial index).

\begin{align}
    \bold{U}_{j+1} = \bold{U}_{j} + \sum_{i=1}^{k}\frac{\Delta t}{\Delta x_{i}}\Bigg[G_{i}(\bold{U}_{j}^{l_{i,1}},\dots,\bold{U}_{j}^{l_{i,m}};\theta_{i})-G_{i}(\bold{U}_{j}^{r_{i,1}},\dots,\bold{U}_{j}^{r_{i,m}};\theta_{i})\Bigg]. \tag{6} \label{eq:6}
\end{align} 
\noindent
To train the Flux NO, the required training dataset shape is [batch size, $N_{t}$, $N_{x_{1}}$, $\cdots$, $N_{x_{k}}$, $N_{u}$] where $N_{t}$ represents the number of interations along time, $N_{x_{i}}$ represent the number of grid points along each spatial dimensions across which the problem is defined, and $N_{u}$ indicates the dimensionality of the problem (e.g., velocity components in fluid dynamics). Given this structured dataset, the loss function motivated from equation \eqref{eq:6} for training Flux NO can be constructed as follows:

\begin{align}
    \mathcal{L}_{tm}(\{\bold{U}_{i}\}_{i=1}^{B})&:=\sum_{i=1}^{B}\sum_{j=1}^{N_{t}-1}\Bigg\|(\bold{U}_{i,j+1} - \bold{U}_{i,j}) \nonumber \\&-\Bigg(\sum_{i=1}^{k}\frac{\Delta t}{\Delta x_{i}}\Bigg[G_{i}(\bold{U}_{i,j}^{l_{i,1}},\dots,\bold{U}_{i,j}^{l_{i,m}};\theta_{i})-G_{i}(\bold{U}_{i,j}^{r_{i,1}},\dots,\bold{U}_{i,j}^{r_{i,m}};\theta_{i})\Bigg]\Bigg)\Bigg\|^{2}.  \tag{7} \label{eq:7}
\end{align} 
This loss function quantifies the difference between residual constructed from the predicted flux values by the Flux NO and the actual (or target) residual values, aiming to minimize this discrepancy during the training process. By focusing on flux approximation, the Flux NO offers a detailed and localized understanding of the underlying physical processes, making it especially suitable for problems governed by conservation laws where the flux plays a critical role. And we consider additional loss which guarantees the consistency of Flux function which is essential for convergence to weak solutions:

\begin{align}
\mathcal{L}_{consi}(\{\bold{U}_{i}\}_{j=1}^{B}):=\sum_{i=1}^{B}\sum_{j=1}^{N^{t}}\sum_{l=1}^{k}\|G_{l}(\bold{U}_{i,j},\dots,\bold{U}_{i,j};\theta_{l})-F_{l}(\bold{U}_{i,j})\|^{2}.   \tag{8} \label{eq:8}
\end{align}  
\noindent
where $F_{i}$ are actual physical fluxes. The actual training is implemented by optimizing weighted sum of these losses: $\mathcal{L}_{tm}+\lambda\mathcal{L}_{consi}$. These losses are presented also in (\cite{Kim:24}), in this work, we consider more additional losses which make the convergnce of  training well and approximate propery with robustness and more generalization ability.

}
\section{Methods}
\subsection{Adaptive FNO Architecture for Multidimensional Outputs in Ideal MHD}{
In the referenced study (\cite{Kim:24}), the output dimension was one-dimensional since it dealt with one-dimensional scalar conservation laws. However, in the case of ideal MHD discussed in this paper, the output dimension is seven-dimensional for one-dimensional problems and eight-dimensional for two-dimensional problems. We experimentally observed performance degradation when a single FNO model was tasked with handling all outputs in cases of large output dimensions; the results of these experiments can be found in Table \ref{t7}. To address this issue, we allocated separate FNO models to handle each physical quantity. Specifically, density, energy, velocity vectors, and magnetic field vectors were each processed by their own FNO models. For two-dimensional problems, given that the physical flux functions exist for both the x and y axes, we further segregated the considerations for each physical flux. This architectural approach is summarized in the schematic shown in the Figure \ref{FLUXMHDNO}. Modified version of \eqref{eq:6} for one-dimensional case can be written as follows:

\begin{align*}
    \bold{U}_{j+1} &= \bold{U}_{j} + \frac{\Delta t}{\Delta x}\Big[G^{l}-G^{r}\Big]  \\
    G^{l}&=\Big(G_{\rho}(\bold{U}_{j}^{l_{1}},\dots,\bold{U}_{j}^{l_{m}};\theta_{\rho}), 
    G_{\bold{u}}(\bold{U}_{j}^{l_{1}},\dots,\bold{U}_{j}^{l_{m}};\theta_{\bold{u}})^{T},  \\
    &G_{\bold{B}}(\bold{U}_{j}^{l_{1}},\dots,\bold{U}_{j}^{l_{m}};\theta_{\bold{B}})^{T}, 
    G_{E}(\bold{U}_{j}^{l_{1}},\dots,\bold{U}_{j}^{l_{m}};\theta_{E})\Big)^{T}   \tag{9} \label{eq:9}  \\
    G^{r}&=\Big(G_{\rho}(\bold{U}_{j}^{r_{1}},\dots,\bold{U}_{j}^{r_{m}};\theta_{\rho}), 
    G_{\bold{u}}(\bold{U}_{j}^{r_{1}},\dots,\bold{U}_{j}^{r_{m}};\theta_{\bold{u}})^{T},  \\
    &G_{\bold{B}}(\bold{U}_{j}^{r_{1}},\dots,\bold{U}_{j}^{r_{m}};\theta_{\bold{B}})^{T}, 
    G_{E}(\bold{U}_{j}^{r_{1}},\dots,\bold{U}_{j}^{r_{m}};\theta_{E})\Big)^{T}.
\end{align*}
\noindent
And the two-dimensional case can be written as follows:

\begin{align*}
    \bold{U}_{j+1} &= \bold{U}_{j} + \frac{\Delta t}{\Delta x}\Big[G^{l}-G^{r}\Big] + \frac{\Delta t}{\Delta y}\Big[F^{t}-F^{b}\Big]  \\
    G^{l}&=\Big(G_{\rho}(\bold{U}_{j}^{l_{1}},\dots,\bold{U}_{j}^{l_{m}};\theta_{\rho}^{G}), 
    G_{\bold{u}}(\bold{U}_{j}^{l_{1}},\dots,\bold{U}_{j}^{l_{m}};\theta_{\bold{u}}^{G})^{T},  \\
    &G_{\bold{B}}(\bold{U}_{j}^{l_{1}},\dots,\bold{U}_{j}^{l_{m}};\theta_{\bold{B}}^{G})^{T}, 
    G_{E}(\bold{U}_{j}^{l_{1}},\dots,\bold{U}_{j}^{l_{m}};\theta_{E}^{G})\Big)^{T} \\
    G^{r}&=\Big(G_{\rho}(\bold{U}_{j}^{r_{1}},\dots,\bold{U}_{j}^{r_{m}};\theta_{\rho}^{G}), 
    G_{\bold{u}}(\bold{U}_{j}^{r_{1}},\dots,\bold{U}_{j}^{r_{m}};\theta_{\bold{u}}^{G})^{T},  \\
    &G_{\bold{B}}(\bold{U}_{j}^{r_{1}},\dots,\bold{U}_{j}^{r_{m}};\theta_{\bold{B}}^{G})^{T}, 
    G_{E}(\bold{U}_{j}^{r_{1}},\dots,\bold{U}_{j}^{r_{m}};\theta_{E}^{G})\Big)^{T}  \tag{10} \label{eq:10} \\
    F^{t}&=\Big(F_{\rho}(\bold{U}_{j}^{t_{1}},\dots,\bold{U}_{j}^{t_{m}};\theta_{\rho}^{F}), 
    F_{\bold{u}}(\bold{U}_{j}^{t_{1}},\dots,\bold{U}_{j}^{t_{m}};\theta_{\bold{u}}^{F})^{T},  \\
    &F_{\bold{B}}(\bold{U}_{j}^{t_{1}},\dots,\bold{U}_{j}^{t_{m}};\theta_{\bold{B}}^{F})^{T}, 
    F_{E}(\bold{U}_{j}^{t_{1}},\dots,\bold{U}_{j}^{t_{m}};\theta_{E}^{F})\Big)^{T} \\
    F^{b}&=\Big(G_{\rho}(\bold{U}_{j}^{b_{1}},\dots,\bold{U}_{j}^{b_{m}};\theta_{\rho}^{F}), 
    F_{\bold{u}}(\bold{U}_{j}^{b_{1}},\dots,\bold{U}_{j}^{b_{m}};\theta_{\bold{u}}^{F})^{T},  \\
    &F_{\bold{B}}(\bold{U}_{j}^{b_{1}},\dots,\bold{U}_{j}^{b_{m}};\theta_{\bold{B}}^{F})^{T}, 
    F_{E}(\bold{U}_{j}^{b_{1}},\dots,\bold{U}_{j}^{b_{m}};\theta_{E}^{F})\Big)^{T}.
\end{align*} 
\noindent
Let the residual in \eqref{eq:9} and \eqref{eq:10} be $\Delta \bold{R}(\bold{U}_{j};\theta):=\bold{U}_{j+1}-\bold{U}_{j}$ where $\theta$ is collection of parameters. Then the loss \eqref{eq:7} can be now written as follows:

\begin{align}
    \mathcal{L}_{tm}(\{\bold{U}_{i}\}_{i=1}^{B})&:=\sum_{i=1}^{B}\sum_{j=1}^{N_{t}-1}\Big\|(\bold{U}_{i,j+1} - \bold{U}_{i,j}) - \Delta \bold{R}(\bold{U}_{i,j};\theta)\Big\|^{2}. \tag{11} \label{eq:11}
\end{align}

\begin{figure}[htp!]
\centering
\includegraphics[height=15.0cm]{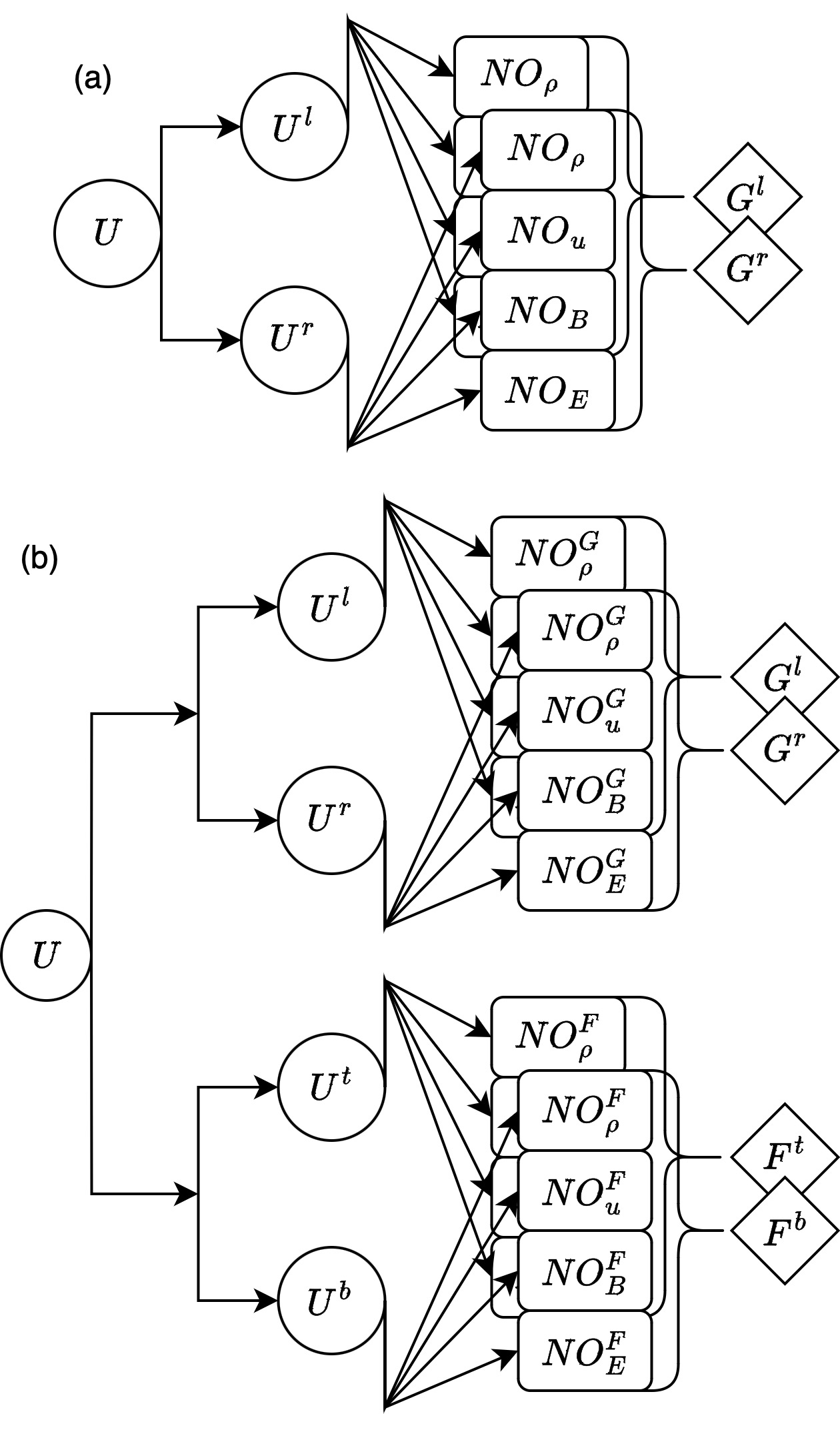}

\caption{Schematic of the calculation of numerical fluxes based on Flux NO models for (a) one-dimensional ideal MHD (b) two-dimensional ideal MHD cases.}\label{FLUXMHDNO}
\end{figure}

}
\noindent
\subsection{Enhanced Loss Function for Ideal MHD}{
We have developed additional loss functions to enhance the existing loss function introduced in (\cite{Kim:24}), and to impart suitable inductive biases for specialized issues such as ideal MHD, which is the focus of this paper. Three further loss functions have been considered; one associated with the Total Variation Diminishing (TVD) property, another related to the divergence-free condition, and a third concerning local information about the flux.

\noindent
{\bf TVD Loss}{
According to Godunov's theorem, the WENO method, which performs high-order polynomial approximations for all stencils, does not inherently possess the Total Variation Diminishing (TVD) property. Nevertheless, WENO exhibits a considerably non-oscillatory manner. The results we will present in Section 4 are based on models trained using training data generated by the high-order WENO method. Although the training data generated through the WENO method do not satisfy the TVD property, we have designed a loss function that imparts an inductive bias towards the TVD property to our model, inspired by equation \eqref{eq:4}. Since the training data inherently lack the TVD property, we adopted an $l^{2}$ rather than $l^{1}$ function as a loss to apply the inductive bias in a soft manner. The specific formulation of this loss is as follows:

\begin{flalign*}
\mathcal{L}_{TVD}(\{\bold{U}_{i}\}_{i=1}^{B}):=\sum_{i=1}^{B}\sum_{n=1}^{N_{t}-1}\lfloor TV(\bold{\tilde{U}}_{i,n+1})-TV(\bold{U}_{i,n})\rfloor_{+}^{2}.   \tag{12} \label{eq:12}
\end{flalign*}
where $\lfloor \cdot\rfloor_{+}:=max(0,x)$ and $\bold{\tilde{U}}$ is an output of Flux NO.
}

\noindent
{\bf Divergence Free Loss}{ Since the magnetic vector field in multidimensional ideal MHD is divergence-free, our approximated fluxes should also construct a divergence-free vector field. To achieve this, we have designed a loss function that imparts this condition. This approach can also be applied to other problems, such as incompressible fluids, which require a divergence-free condition for the velocity vector field.
Each of $\bold{F}$ and $\bold{G}$ represents the physical fluxes for the two-dimensional ideal MHD equations, as introduced in Section 2.1. The subscript $\bold{B}$ in \eqref{eq:13} indicates that the components of the vector are specifically related to the magnetic vector fields.

\begin{align}
&\nabla \cdot \frac{\partial\bold{U}}{\partial t}=  \frac{\partial \nabla \cdot \bold{U}}{\partial t} = \nabla \cdot \Big(-\frac{\partial \bold{G}}{\partial x} -\frac{\partial \bold{F}}{\partial y}\Big) \nonumber \\
&\Rightarrow 0= \frac{\partial \nabla \cdot \bold{B}}{\partial t}  = \nabla \cdot \Big(-\frac{\partial \bold{G}}{\partial x} -\frac{\partial \bold{F}}{\partial y}\Big)_{\bold{B}}. \tag{13} \label{eq:13}
\end{align} 

\noindent
According to equation \eqref{eq:13}, the residual of the magnetic field must also satisfy the divergence-free condition. Therefore, we will design a loss function to ensure that the residuals constructed from our approximated flux satisfy this condition. However, the training datasets generated through numerical analysis may still exhibit non-zero divergence of the magnetic field, despite the application of divergence-free relaxation. Consequently, instead of implementing a hard constraint that forces the divergence-related loss to converge nearly to zero, or requiring the model architecture to output the residual of the magnetic field as the curl of a specific vector field, we empirically calculate the average or maximum values of the magnetic field's divergence in the training dataset. Using this information, we set a threshold such that the $l^{2}$ loss related to the divergence is applied only when the model's output exceeds this threshold. The specific mathematical formulation of this loss is as follows:

\begin{flalign*}
\mathcal{L}_{div}(\{\bold{U_{i}}\}_{i=1}^{B}):=\sum_{i=1}^{B}\sum_{j=1}^{N_{t}-1} \Bigg\lfloor \frac{\|\nabla\cdot\Delta \bold{R}(\bold{U}_{i,j};\theta)_{\bold{B}}\|_{2}^{2}-\theta_{div}}{| \|\nabla\cdot\Delta \bold{R}(\bold{U}_{i,j};\theta)_{\bold{B}}\|_{2}^{2}-\theta_{div}|}\Bigg\rfloor_{+}\|\nabla\cdot\Delta \bold{R}(\bold{U}_{i,j};\theta)_{\bold{B}} \|_{2}^{2}.   \tag{14} \label{eq:14}
\end{flalign*}
where $\theta_{div}$ is the threshold value.
}

\noindent
{\bf $l^{\infty}$ Loss}{
In addition to the $l^{2}$ norm of the residuals proposed in (\cite{Kim:24}), we also consider the $l^{\infty}$ norm, which is used to handle outlier values (\cite{Bar:21}). Utilizing the $l^{\infty}$ norm allows for the expectation of pointwise convergence, thereby enabling a more accurate approximation of the flux embedded in the training dataset. In cases like ideal MHD where the output is vector-valued, we apply the $l^{\infty}$ norm to each vector component and then sum these values. The specific mathematical expression for this is as follows:
\begin{flalign*}
\mathcal{L}_{\infty}(\{\bold{U_{i}}\}_{i=1}^{B}):=\sum_{i=1}^{B}\sum_{n=1}^{N_{t}-1}\sum_{l=1}^{N_{u}}\sup_{j,k} \Big((\bold{U}_{i,n+1,j,k,l} - \bold{U}_{i,n,j,k,l}) - \Delta \bold{R}(\bold{U}_{i,n,j,k,l};\theta)\Big).   \tag{15} \label{eq:15}
\end{flalign*}
}

}

\subsection{Overall Algorithm}{ 
Now, by combining the defined loss functions in Equations \eqref{eq:8}, \eqref{eq:11}, \eqref{eq:12}, \eqref{eq:14}, and \eqref{eq:15}, we demonstrate the training algorithm (Algorithm \ref{alg:1}) for our modified Flux Neural Operator (Flux NO). Since inference is straightforwardly conducted using Equation \eqref{eq:6}, we omit the details. We focus on the two-dimensional case, as the training process for the one-dimensional case is similar to the algorithm proposed in (\cite{Kim:24}), with the exception of the enhanced loss function.

\begin{algorithm}
\caption{An algorithm for training}\label{alg:1}
\hspace*{\algorithmicindent} \textbf{Input:}{\quad Dataset $\mathcal{U}=((U_{\tilde{b},i,j,k,m}),(\Delta t_{b,i}))$}  \\
\hspace*{\algorithmicindent} \textbf{Output:}{\quad trained FNO models $G_{\rho}(\cdot;\theta_{\rho}^{G})$, $G_{\bold{u}}(\cdot;\theta_{\bold{u}}^{G})$, $G_{\bold{B}}(\cdot;\theta_{\bold{B}}^{G})$, $G_{E}(\cdot;\theta_{E}^{G})$ and $F_{\rho}(\cdot;\theta_{\rho}^{F})$, $F_{\bold{u}}(\cdot;\theta_{\bold{u}}^{F})$, $F_{\bold{B}}(\cdot;\theta_{\bold{B}}^{F})$, $F_{E}(\cdot;\theta_{E}^{F})$}
\begin{algorithmic}
\For{\texttt{epoch$=1,\dots,E$}}
\For{\texttt{Batch $\in$ Train loader}}
    \State $\tilde{U}_{-\tilde{j}} \gets$ roll $U_{\tilde{b},i,\cdot,\cdot,m}$ by $\tilde{j}$ in the third index for $\tilde{j} = -q,\dots,p+1$  
    \State $\tilde{U}^{-\tilde{k}} \gets$ roll $U_{\tilde{b},i,\cdot,\cdot,m}$ by $\tilde{k}$ in the fourth index for $\tilde{k} = -q,\dots,p+1$  
    \State $U^{l} \gets$ concatenate $(\tilde{U}_{-p},\dots,\tilde{U}_{q})$ along fifth index
    \State $U^{r} \gets$ concatenate $(\tilde{U}_{-p-1},\dots,\tilde{U}_{q-1})$ along fifth index
    \State $U^{t} \gets$ concatenate $(\tilde{U}^{-p},\dots,\tilde{U}^{q})$ along fifth index 
    \State $U^{b} \gets$ concatenate $(\tilde{U}^{-p-1},\dots,\tilde{U}^{q-1})$ along fifth index \Comment{Thus, the concatenated function is now a 8(p+q)-dimensional vector-valued function}
    \State $G^{l} \gets$ concatenate $(G_{\rho}(U^{l};\theta_{\rho}^{G}),G_{\bold{u}}(U^{l};\theta_{\bold{u}}^{G}),G_{\bold{B}}(U^{l};\theta_{\bold{B}}^{G}),G_{E}(U^{l};\theta_{E}^{G}))$ along fifth index 
    \State $G^{r} \gets$ concatenate $(G_{\rho}(U^{r};\theta_{\rho}^{G}),G_{\bold{u}}(U^{r};\theta_{\bold{u}}^{G}),G_{\bold{B}}(U^{r};\theta_{\bold{B}}^{G}),G_{E}(U^{r};\theta_{E}^{G}))$ along fifth index 
    \State $F^{t} \gets$ concatenate $(F_{\rho}(U^{t};\theta_{\rho}^{F}),F_{\bold{u}}(U^{t};\theta_{\bold{u}}^{F}),F_{\bold{B}}(U^{t};\theta_{\bold{B}}^{F}),F_{E}(U^{t};\theta_{E}^{F}))$ along fifth index 
    \State $F^{b} \gets$ concatenate $(F_{\rho}(U^{b};\theta_{\rho}^{F}),F_{\bold{u}}(U^{b};\theta_{\bold{u}}^{F}),F_{\bold{B}}(U^{b};\theta_{\bold{B}}^{F}),F_{E}(U^{b};\theta_{E}^{F}))$ along fifth index 
    \State $\Delta \bold{R}(\bold{U}_{\tilde{b},i,\cdot,\cdot,\cdot};\theta) \gets$  $ \frac{\Delta t_{\tilde{b},i}}{\Delta x}\Big[G^{l}-G^{r}\Big] + \frac{\Delta t_{\tilde{b},i}}{\Delta y}\Big[F^{t}-F^{b}\Big] $
    \State $\mathcal{L}_{tm}(Batch) \gets \sum_{\tilde{b}=1}^{B}\sum_{i=1}^{N_{t}-1}\|U_{\tilde{b},i+1,\cdot,\cdot,\cdot}-U_{\tilde{b},i,\cdot,\cdot,\cdot}-\Delta\bold{R}(\bold{U}_{\tilde{b},i,\cdot,\cdot,\cdot};\theta)\|_{2}^{2}$ \Comment{B is batch size}
    \State $V^{p+q} \gets$ concatenate $U$ p+q times.
    \State $\mathcal{L}_{consi}(Batch) \gets \sum_{\tilde{b}=1}^{B}\sum_{i=1}^{N_{t}}\Big(\|\tilde{G}(V^{p+q}_{\tilde{b},i,\cdot,\cdot,\cdot};\theta)-G(V^{p+q}_{\tilde{b},i,\cdot,\cdot,\cdot})\|_{2}^{2}+\|\tilde{F}(V^{p+q}_{\tilde{b},i,\cdot,\cdot,\cdot};\theta)-F(V^{p+q}_{\tilde{b},i,\cdot,\cdot,\cdot})\|_{2}^{2}\Big)$  \Comment {Each of $\tilde{G}$ and $\tilde{F}$ represents concatenated numerical fluxes, and $G$ and $F$ are the physical fluxes.}
    \State $\mathcal{L}_{TVD}(Batch) \gets \sum_{\tilde{b}=1}^{B}\sum_{i=1}^{N_{t}-1}\|\lfloor TV(U_{\tilde{b},i,\cdot,\cdot,\cdot}+\Delta\bold{R}(\bold{U}_{\tilde{b},i,\cdot,\cdot,\cdot};\theta))-TV(U_{\tilde{b},i,\cdot,\cdot,\cdot})\rfloor_{+}\|_{2}^{2}$ 
    \State $\mathcal{L}_{\infty}(Batch) \gets \sum_{\tilde{b}=1}^{B}\sum_{i=1}^{N_{t}-1}\sum_{l=1}^{N_{u}}\sup_{j,k}\|U_{\tilde{b},i+1,j,k,l}-U_{\tilde{b},i,j,k,l}-\Delta\bold{R}(\bold{U}_{\tilde{b},i,j,k,l};\theta)\|_{2}^{2}$ 
    \State $ \mathcal{L}_{div}(Batch) \gets\sum_{\tilde{b}=1}^{B}\sum_{i=1}^{N_{t}-1} \Bigg\lfloor \frac{\|\nabla\cdot\Delta \bold{R}(\bold{U}_{\tilde{b},i};\theta)_{\bold{B}}\|_{2}^{2}-\theta_{div}}{| \|\nabla\cdot\Delta \bold{R}(\bold{U}_{\tilde{b},i};\theta)_{\bold{B}}\|_{2}^{2}-\theta_{div}|}\Bigg\rfloor_{+}\|\nabla\cdot\Delta \bold{R}(\bold{U}_{\tilde{b},i};\theta)_{\bold{B}} \|_{2}^{2}     $
    \State Calculate backpropagation for \\ $\lambda_{tm}\mathcal{L}_{tm}(Batch)+\lambda_{consi}\mathcal{L}_{consi}(Batch)+\lambda_{TVD}\mathcal{L}_{TVD}(Batch)+\lambda_{\infty}\mathcal{L}_{\infty}(Batch)+\lambda_{div}\mathcal{L}_{div}(Batch)$ \Comment{Each $\lambda$ with a subscript is a weight for the losses.}
    \State Take an optimization step.
\EndFor
\EndFor
\end{algorithmic}
\end{algorithm}

}

\section{Results} 
\subsection{Training Dataset and Architecture of Flux Neural Operator}{
{\bf One-dimensional case}{
\noindent
For the one-dimensional problem, we constructed the initial conditions using Gaussian random fields. Specifically, we set the initial values for density and the $B_{x}$ to 1, and specified the velocity and magnetic field components as vectors discretized to size 256, derived from Gaussian random fields with a covariance of $k(x,y)=e^{-100(x-y)^{2}}$. To ensure the well-posedness of the problem, each component of the velocity and magnetic field was scaled by 0.3. Given that the $\gamma$ value was set to 2.0, the initial value of the energy is determined by the following equation: $E=1.0+\frac{1}{2}\rho\bold{u}^{2}+\frac{1}{2}\|\bold{B}\|^{2}$. Based on these initial conditions, we solved the problem using the WENO-JS scheme for spatial order 5 and the RK method of order 4 for temporal discretization. The CFL number was 0.3. Based on this scheme, we divided the solution from $t=0.0$ to $t=0.6$ into 600 segments and then extracted snapshots only from $t=0.1$ to $t=0.6$ to form a dataset consisting of 500 snapshots for each function. We generated 500 tensors representing functions in $C^{\infty}([0.1,0.6]\times[0,1])$ to create the training dataset. In the model architecture, the Neural Operator component fundamentally employs the FNO; however, as FNO performs poorly with non-periodic functions, all datasets were composed as periodic functions. We anticipate finding suitable models for non-periodic functions as well, since the Neural Operator component can be replaced with other models. For the one-dimensional problem discussed in this paper, the model’s architecture uses 5 modes, width 64, and depth 3, with output dimensions of 1, 3, 2, and 1 for density, velocity, magnetic field, and energy, respectively. The kernel size of the CNN layer in the Fourier layer is uniformly set to 1.

}

\noindent
{\bf Two-dimensional case}{
\noindent
For the two-dimensional problem, initial conditions were generated using two-dimensional Gaussian random fields. We independently sampled the components of velocity and magnetic field vectors from Gaussian random fields. And set the initial density value to $\gamma^{2}$. The Gaussian random fields we used have a power spectrum $P(k)\propto k^{-2.5}$. Unlike the one-dimensional case, in two dimensions, the divergence of the magnetic field must be zero for physically meaningful results and for the numerical scheme to be stable; therefore, the randomly sampled magnetic fields were relaxed using the Poisson equation (\cite{Jiang:99}). The grid size for the initial conditions was set to 64 along both the x and y axes, and the $\gamma$ value was set to 5/3. The initial energy value will be determined by the following equation: $E=\frac{5}{2}+\frac{1}{2}\rho\bold{u}^{2}+\frac{1}{2}\|\bold{B}\|^{2}$. Based on these initial conditions, we used the WENO-Z scheme for spatial order 5, temporal discretization of fourth order RK method, and set the CFL numbers to 0.4. Under this scheme, we divided the solution from  $t=0.0$ to $t=0.75$ into 150 segments, and then extracted snapshots only from $t=0.25$ to $t=0.75$ to form a dataset consisting of 100 snapshots per function. In this way, we created 100 tensors, each representing a function in $C^{\infty}([0.25,0.75]\times[0,2\pi]^{2})$ for training. The model architecture adopted a 2D FNO for the neural operator component, with architecture hyperparameters including 4 modes each for the x and y axes, a width of 72, and a depth of 2. These hyperparameters were empirically optimized within the resource constraints available to us. Increasing the maximal frequency did not result in significant performance improvements, which is presumed to be because the physics of ideal MHD inherently propagates information locally. The output dimensions for models concerning density, velocity, magnetic field, and energy are 1, 3, 3, and 1, respectively. The kernel size of the CNN layer in the Fourier layer is uniformly set to 1. Both one-dimensional and two-dimensional problems use the GELU activation function. Each neural operator takes concatenated tensors composed of eight tensors, which are shifted circularly from -3 to 4 (where an $i$ shift means $j \rightarrow j+i$) for $G^{l}$ and $F^{t}$, and from -4 to 3 for $G^{r}$ and $F^{b}$ for corresponding indices.

}

}

\subsection{Generalization Ability}{
The results in sections 4.2 to 4.4 were obtained using models trained under the following hyperparameters. For the one-dimensional case: the optimizer was Adam with a learning rate of $1e-3$ and weight decay of $1e-3$. The scheduler was CosineAnnealingWarmRestarts with $T_{0}=100$ and eta\_min=$1e-4$. The batch size was 5, $\lambda_{tm}=1.0$, $\lambda_{TVD}=5e-3$, $\lambda_{\infty}=1.0$, and $\lambda_{consi}=1.0$. For the two-dimensional case: the optimizer was Adam with a learning rate of $4e-4$ and weight decay of $2e-4$. The scheduler was CosineAnnealingWarmRestarts with $T_{0}=100$ and eta\_min=$1e-5$. The batch size was 1, $\lambda_{tm}=4.0$, $\lambda_{TVD}=2.5e-4$, $\lambda_{div}=1e-2$, $\lambda_{\infty}=1.0$, and $\theta_{div}=150$. Where terms such as $T_{0}$ and eta\_min follow the conventions used in PyTorch. In this section, we analyze the results of inferences made by our model on examples that are within the test distribution but were not utilized as training data, both qualitatively and quantitatively. Additionally, we experimentally demonstrate that our model can make inferences over longer time periods than those covered by the functions used in the training dataset for the test samples, and we also analyze results under conditions of higher resolution.

\noindent
{\bf Short term inference}{
\noindent
We conducted experiments on test samples that were sampled from the same distribution as the training dataset and had the same short-term span lengths as the training samples ($t=0.1$ to $0.6$ for the one-dimensional case, $t=0.25$ to $0.75$ for the two-dimensional case). The results for the one-dimensional case are presented in Figure \ref{1D_short} and Table \ref{t1}, while the results for the two-dimensional case are shown in Figure \ref{2D_short_rho_E}, \ref{2D_short_u_B}, \ref{2D_short_Error}, \ref{2D_short_section} and Table \ref{t2}. Tables \ref{t1} and \ref{t2} provide statistics on the relative $l^{2}$ norm and $l^{\infty}$ norm across 10 test samples. Due to the smaller domain size in the one-dimensional case compared to the two-dimensional case, the changes in the solutions are observed to be more dynamic. In addition, in the one-dimensional case, the solution generated by Flux NO tends to lose stability more easily because the grid size of 256 (compared to 64 in each axis for the two-dimensional case) makes it easier to violate the CFL condition. For the one-dimensional problem, the instability of Flux NO is such that only results up to $t=0.2$ are shown in Table \ref{t1}. From Table \ref{t1}, it can be seen that the norms related to velocity are the largest and increase the fastest, which can be attributed to the poorer approximation of the $u_{y}$ and $u_{z}$ components, while $u_{x}$ is well approximated compared to these, as indicated by Figure \ref{1D_short}. Consequently, although the training dataset for the one-dimensional problem is richer, consisting of 500 functions compared to 100 for the two-dimensional problem, it is expected to have lower stability and generalization power due to the reasons mentioned above. For the two-dimensional problem, the performance on test samples within the short-term training distribution appears to be satisfactory, as can be seen from Figure \ref{2D_short_rho_E}, \ref{2D_short_u_B}, \ref{2D_short_Error}, \ref{2D_short_section} and Table \ref{t2}, indicating that the solution for each component has been well resolved.

\begin{figure}
  \makebox[\textwidth][c]{\includegraphics[width=1.3\textwidth]{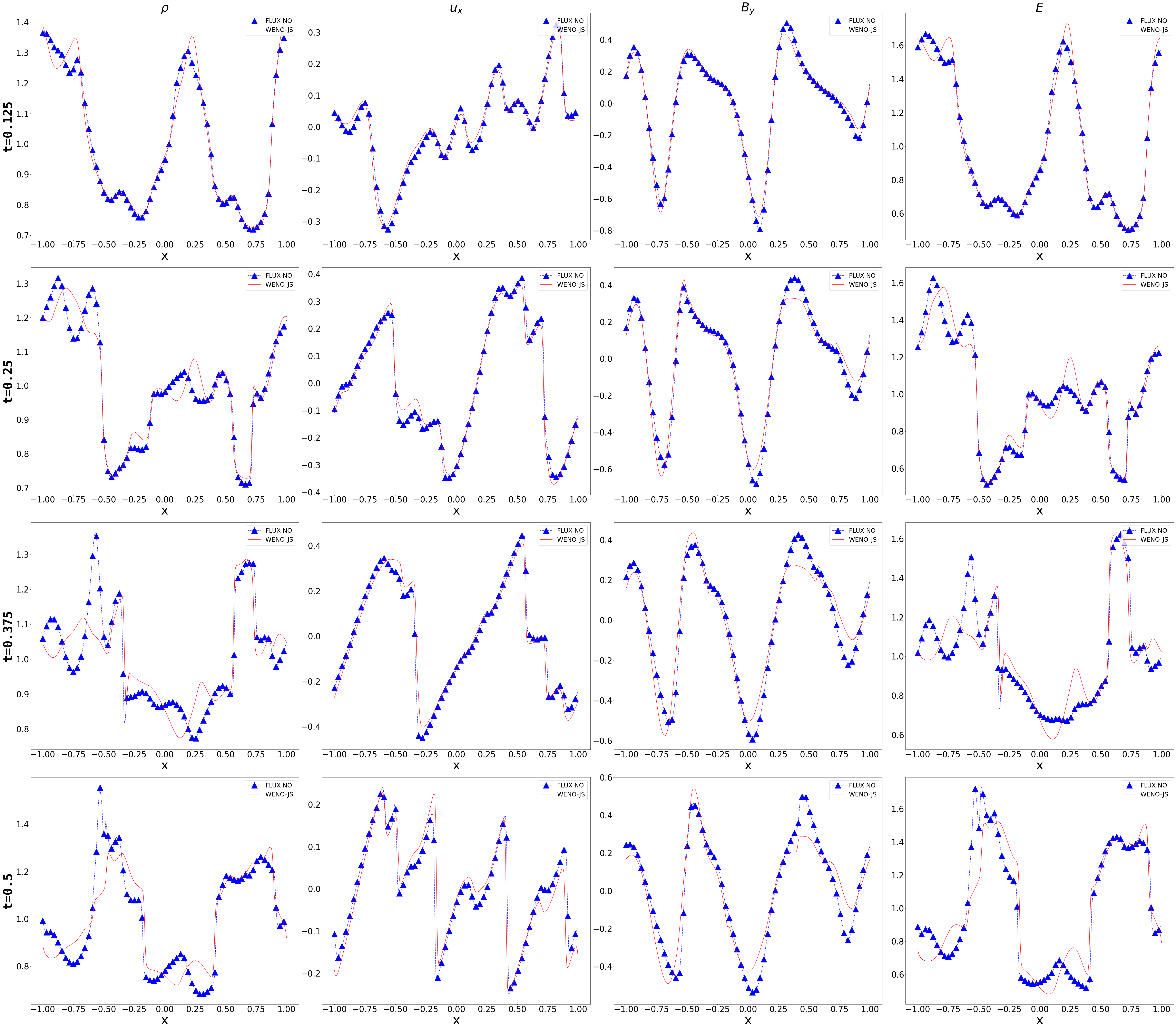}}%
  \caption{Time Evolution of Each Component of the Output from Flux NO and Reference Data for One-Dimensional Ideal MHD.}
  \label{1D_short}
\end{figure}

\begin{table}[H]
\centering
\begin{tabular}{lllll}
\hline\noalign{\smallskip}
\thead{(M, SD)} & \thead{t=0.05} & \thead{t=0.1} & \thead{t=0.15} & \thead{t=0.2} \\
\hline
\thead{rel $l^{2}_\rho$} & (3e-4, 1.10e-1) & (1e-3, 1.10e-1) & (1.9e-3, 1.11e-1) & (3.5e-3, 1.11e-1)\\ 
\hline
\thead{rel $l^{\infty}_\rho$} & (4.08e-8, 1.08e-1) & (4.76e-7, 1.09e-1) & (1.97e-6, 1.09e-1) & (4.77e-6, 1.09e-1)\\ 
\hline
\thead{rel $l^{2}_\bold{u}$}  & (7.5e-3, 1.11e-1)  & (1.47e-2, 1.11e-1) & (2.4e-2, 1.11e-1) & (3.52e-2, 1.11e-1)\\ 
\hline
\thead{rel $l^{\infty}_\bold{u}$}  & (6e-4, 1.10e-1)  & (4.1e-3, 1.11e-1) & (1.07e-2, 1.11e-1) & (2.23e-2, 1.11e-1)\\ 
\hline
\thead{rel $l^{2}_\bold{B}$}  & (4.5e-3, 1.11e-1)  & (8.4e-3, 1.11e-1) & (1.15e-2, 1.11e-1) & (1.51e-2, 1.11e-1) \\
\hline
\thead{rel $l^{\infty}_\bold{B}$}  & (2e-4, 1.10e-1)  & (1.9e-3, 1.11e-1) & (2.2e-3, 1.11e-1) & (4.5e-3, 1.11e-1) \\
\hline
\thead{rel $l^{2}_E$} & (8e-4, 1.10e-1)  & (1.9e-3, 1.11e-1) & (3.2e-3, 1.11e-1) & (5e-3, 1.11e-1) \\
\hline
\thead{rel $l^{\infty}_E$} & (2.05e-5, 1.10e-1)  & (3e-4, 1.10e-1) & (4e-4, 1.10e-1) & (2.9e-3, 1.11e-1) \\
\hline

\end{tabular}
\caption{Means (M) and Standard Deviations (SD) of Relative $l^{2}$ and $l^{\infty}$ Norms Between the Output of Flux NO and Reference for Each Component at Various Times in the One-Dimensional Case, Across 10 Test Samples.} \label{t1}
\end{table}

\begin{figure}
  \makebox[\textwidth][c]{\includegraphics[width=1.3\textwidth]{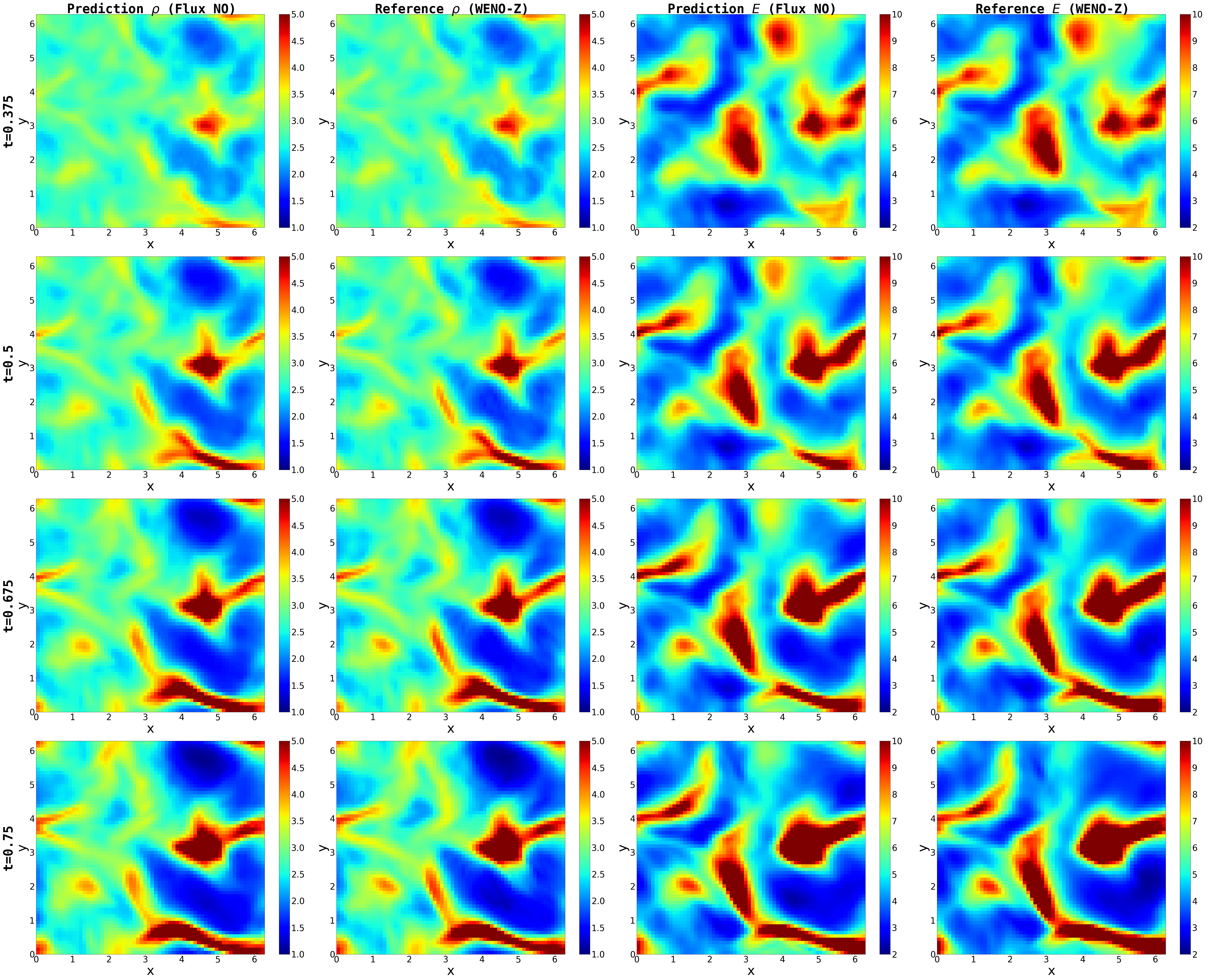}}%
  \caption{Short-Term Time Evolution of $\rho$ and $E$ in the Output from Flux NO Compared to Reference Data for Two-Dimensional Ideal MHD.}
  \label{2D_short_rho_E}
\end{figure}

\begin{figure}
  \makebox[\textwidth][c]{\includegraphics[width=1.3\textwidth]{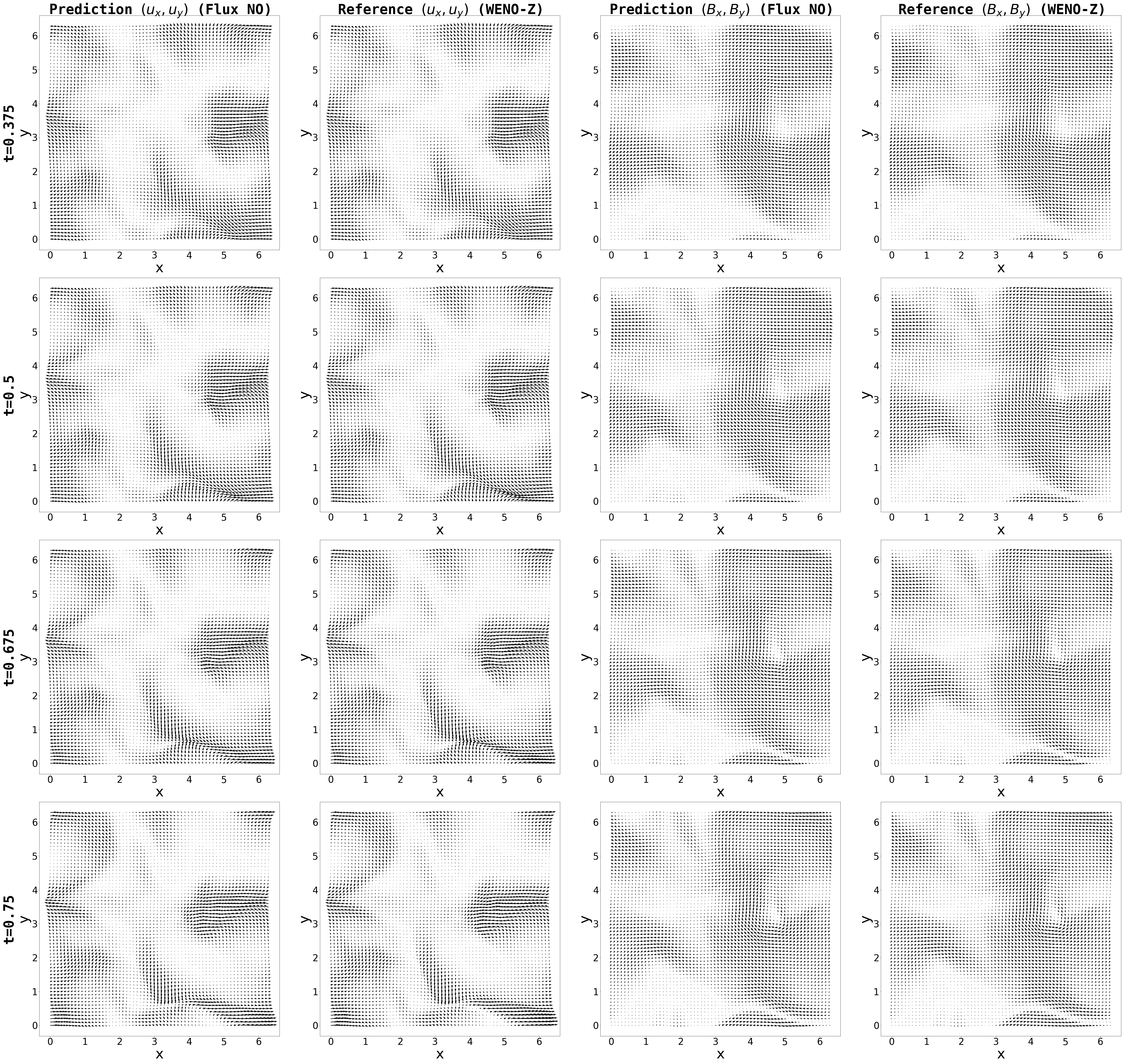}}%
  \caption{Short-Term Time Evolution of $(u_{x},u_{y})$ and $(B_{x},B_{y})$ in the Output from Flux NO Compared to Reference Data for Two-Dimensional Ideal MHD.}
  \label{2D_short_u_B}
\end{figure}

\begin{figure}
  \makebox[\textwidth][c]{\includegraphics[width=1.3\textwidth]{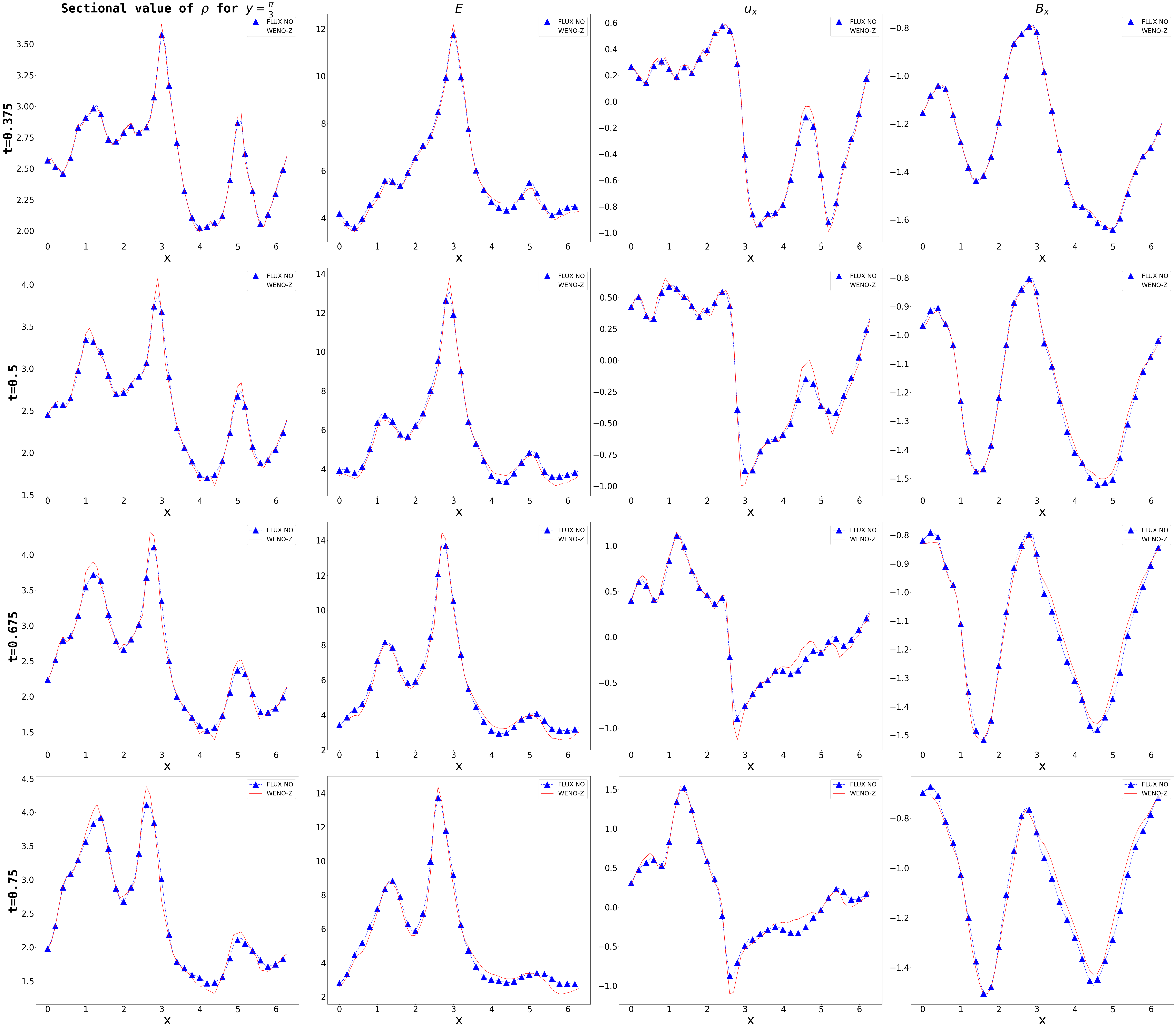}}%
  \caption{Short-Term Time Evolution of Each Component for Section $y=\frac{\pi}{3}$ in the Output from Flux NO Compared to Reference Data for Two-Dimensional Ideal MHD.}
  \label{2D_short_section}
\end{figure}

\begin{table}[H]
\centering
\begin{tabular}{lllll}
\hline\noalign{\smallskip}
\thead{(M, SD)} & \thead{t=0.375} & \thead{t=0.5} & \thead{t=0.675} & \thead{t=0.75} \\
\hline
\thead{rel $l^{2}_\rho$} & (9e-4, 1.10e-1) & (2.5e-3, 1.11e-1) & (4.4e-3, 1.11e-1) & (6.7e-3, 1.11e-1)\\ 
\hline
\thead{rel $l^{\infty}_\rho$} & (5.44e-8, 1.08e-1) & (4.24e-7, 1.09e-1) & (1.52e-6, 1.09e-1) & (5.51e-6, 1.09e-1)\\ 
\hline
\thead{rel $l^{2}_\bold{u}$}  & (2.1e-3, 1.11e-1)  & (4.7e-3, 1.11e-1) & (7.4e-3, 1.11e-1) & (1.02e-2, 1.11e-1)\\ 
\hline
\thead{rel $l^{\infty}_\bold{u}$}  & (9.23e-5, 1.10e-1)  & (3e-4, 1.10e-1) & (6e-4, 1.10e-1) & (1e-3, 1.10e-1)\\ 
\hline
\thead{rel $l^{2}_\bold{B}$}  & (1.6e-3, 1.10e-1)  & (3.5e-3, 1.11e-1) & (5.6e-3, 1.11e-1) & (7.8e-3, 1.11e-1) \\
\hline
\thead{rel $l^{\infty}_\bold{B}$}  & (3.96e-5, 1.10e-1)  & (2e-4, 1.10e-1) & (4e-4, 1.10e-1) & (7e-4, 1.10e-1) \\
\hline
\thead{rel $l^{2}_E$} & (3e-3, 1.11e-1)  & (5.2e-3, 1.11e-1) & (6.8e-3, 1.11e-1) & (8.1e-3, 1.11e-1) \\
\hline
\thead{rel $l^{\infty}_E$} & (5e-4, 1.10e-1)  & (1.6e-3, 1.10e-1) & (3.2e-3, 1.11e-1) & (2.7e-3, 1.11e-1) \\
\hline

\end{tabular}
\caption{Means (M) and Standard Deviations (SD) of Relative $l^{2}$ and $l^{\infty}$ Norms Between the Output of Flux NO and Reference for Each Component at Short Term Times in the Two-Dimensional Case, Across 10 Test Samples.} \label{t2}
\end{table}
}
\noindent
{\bf Long term inference}{
\noindent
We conducted experiments over time spans significantly longer than the time domains of the functions used for training, with the initial conditions configured from the same distribution as training dataset. Due to instability issues, long-term experiments were not conducted for the one-dimensional case; instead, they were carried out only for the two-dimensional case, with the results presented in Figure \ref{2D_long_rho_E}, \ref{2D_long_u_B}, \ref{2D_long_Error}, \ref{2D_long_section} and Table \ref{t3}. The training functions' time domain ranged from $t=0.25$ to $t=0.75$, with a fixed time interval of $\Delta t=0.005$, allowing for the terminal point to be reached after 100 iterations. However, in our results, using the same time intervals, the iterations extended to much longer times: $t=1.25$ (200 iterations), $t=1.5$ (250 iterations), $t=1.75$ (300 iterations), and $t=2.0$ (350 iterations). As shown in Figures \ref{2D_short_Error} and \ref{2D_long_Error}, the error increases almost linearly with the number of iterations, affecting both the global and local characteristics of the solution and leading to progressively increasing discrepancies with the reference, as qualitatively evident in Figures \ref{2D_long_rho_E}, \ref{2D_long_u_B} and \ref{2D_long_section}. A notable aspect of the long-term inference for the two-dimensional problem is that, unlike the one-dimensional case, even with many more iterations, the solutions output by Flux NO did not blow up.

\begin{figure}
  \makebox[\textwidth][c]{\includegraphics[width=1.3\textwidth]{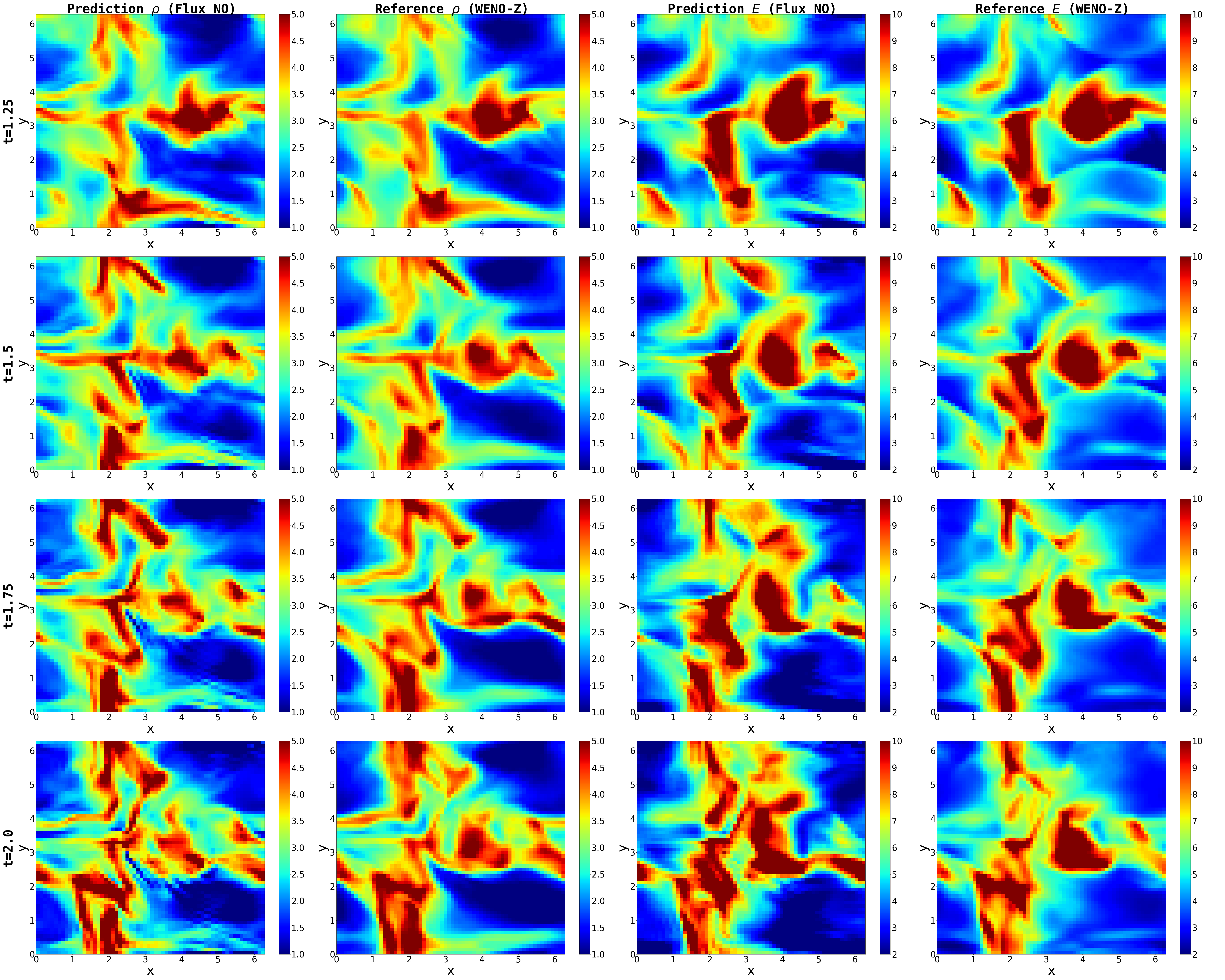}}%
  \caption{Long-Term Time Evolution of $\rho$ and $E$ in the Output from Flux NO Compared to Reference Data for Two-Dimensional Ideal MHD.}
  \label{2D_long_rho_E}
\end{figure}

\begin{table}[H]
\centering
\begin{tabular}{lllll}
\hline\noalign{\smallskip}
\thead{(M, SD)} & \thead{t=1.25} & \thead{t=1.5} & \thead{t=1.75} & \thead{t=2.0} \\
\hline
\thead{rel $l^{2}_\rho$} & (1.31e-2, 1.11e-1) & (1.62e-2, 1.11e-1) & (1.98e-2, 1.11e-1) & (2.45e-2, 1.11e-1)\\ 
\hline
\thead{rel $l^{\infty}_\rho$} & (1.77e-5, 1.10e-1) & (2.78e-5, 1.10e-1) & (2.95e-5, 1.10e-1) & (4.90e-5, 1.10e-1)\\ 
\hline
\thead{rel $l^{2}_\bold{u}$}  & (2.15e-2, 1.11e-1)  & (2.91e-2, 1.11e-1) & (3.86e-2, 1.11e-1) & (4.68e-2, 1.11e-1)\\ 
\hline
\thead{rel $l^{\infty}_\bold{u}$}  & (4.6e-3, 1.11e-1)  & (6.6e-3, 1.11e-1) & (1.45e-2, 1.11e-1) & (1.98e-2, 1.11e-1)\\ 
\hline
\thead{rel $l^{2}_\bold{B}$}  & (1.75e-2, 1.11e-1)  & (2.39e-2, 1.11e-1) & (2.97e-2, 1.11e-1) & (3.59e-2, 1.11e-1) \\
\hline
\thead{rel $l^{\infty}_\bold{B}$}  & (4.5e-3, 1.11e-1)  & (9.2e-3, 1.11e-1) & (1.07e-2, 1.11e-1) & (1.63e-2, 1.11e-1) \\
\hline
\thead{rel $l^{2}_E$} & (1.33e-2, 1.11e-1)  & (1.69e-2, 1.11e-1) & (2.38e-2, 1.11e-1) & (3.16e-2, 1.11e-1) \\
\hline
\thead{rel $l^{\infty}_E$} & (7.3e-3, 1.11e-1)  & (8.6e-3, 1.11e-1) & (4.27e-2, 1.11e-1) & (4.4e-2, 1.11e-1) \\
\hline

\end{tabular}
\caption{Means (M) and Standard Deviations (SD) of Relative $l^{2}$ and $l^{\infty}$ Norms Between the Output of Flux NO and Reference for Each Component at Long Term Times in the Two-Dimensional Case, Across 10 Test Samples.} \label{t3}
\end{table}

}

\noindent
{\bf High resolution inference}{
\noindent
We conducted tests on samples at a higher resolution (96 by 96) than the resolution of the training data (64 by 64). Initial conditions were sampled from a Gaussian random field with the same power spectrum as the training data, and references were computed using the WENO-Z scheme with each snapshot spaced by a $\Delta t=0.005$. The test results are presented in Figures \ref{2D_high_rho_E}, \ref{2D_high_u_B}, \ref{2D_high_section} and \ref{2D_high_Error}; trends in errors, which showed no significant difference from the original resolution, have been omitted. As seen in Figures \ref{2D_high_rho_E} and \ref{2D_high_u_B}, while Flux NO captures the global characteristics of the solution, local instabilities gradually appear in the snapshots at $t=1.5$ and $t=2.0$. This instability, similar to that observed in the one-dimensional case, appears to arise from the higher resolution and is somewhat related to the CFL condition. Attempts to more easily satisfy the CFL condition by reducing the interval to $\Delta t=0.0025$ for denser inference were made, yet instabilities still formed. From these experimental results, we hypothesize that as the resolution of vectorized functions increases, the domain dimension of their data distribution expands, making it challenging to approximate the data distribution itself, leading to these outcomes.

\begin{figure}
  \makebox[\textwidth][c]{\includegraphics[width=1.3\textwidth]{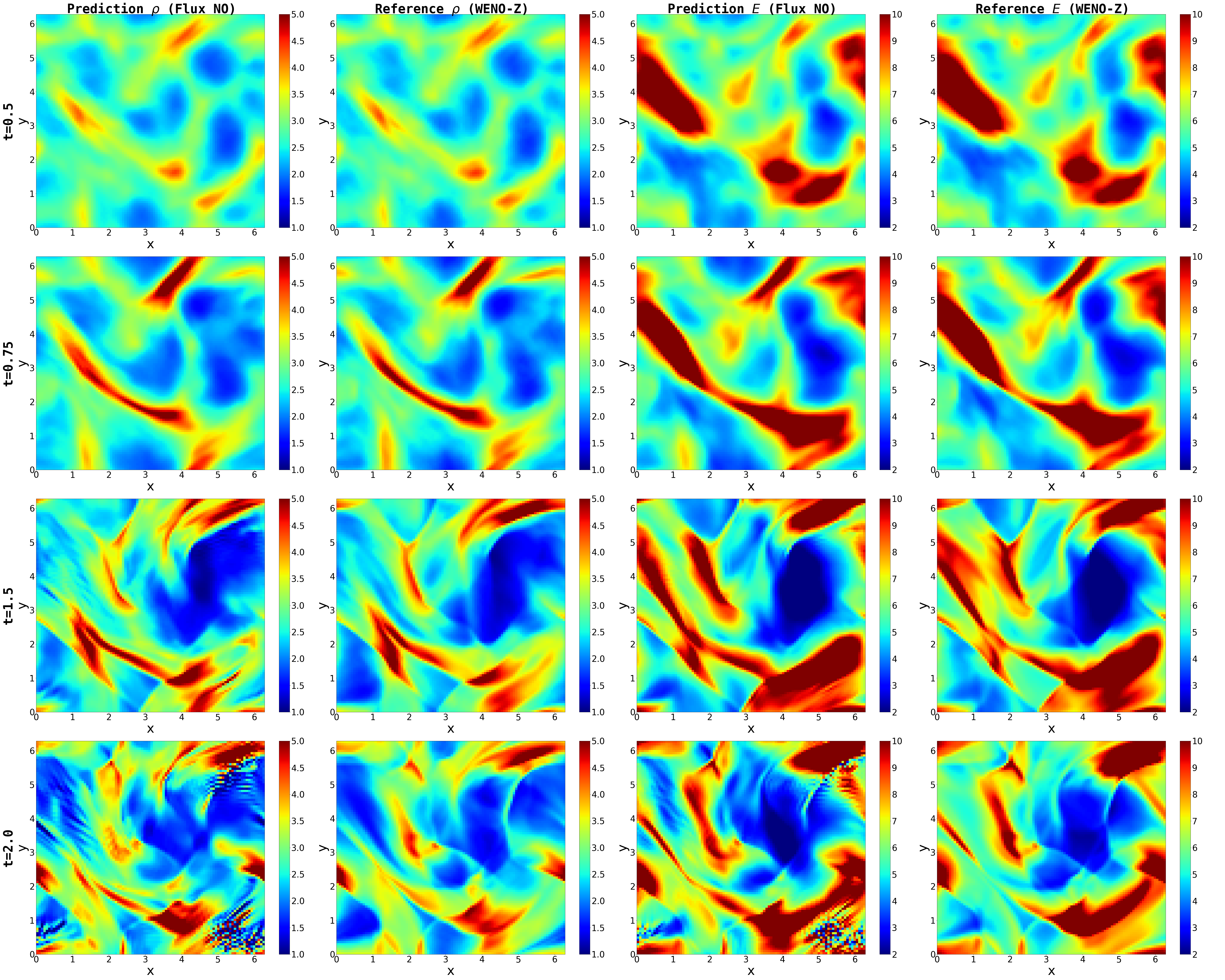}}%
  \caption{Time Evolution of $\rho$ and $E$ in the Output from Flux NO Compared to Reference Data in the Higher Resolution (96 by 96) for Two-Dimensional Ideal MHD.}
  \label{2D_high_rho_E}
\end{figure}

}

}
\subsection{Tests on Out of Distribution Samples}{

\noindent
In this section, we qualitatively evaluate the performance of our model given initial conditions outside the training distribution. We intended to apply the Brio-Wu problem, a type of Riemann problem, but since we assumed periodic boundary conditions for the one-dimensional case, we slightly modified the original Brio-Wu problem and set the initial conditions as follows:
\begin{flalign*}
(\rho,u_{x},u_{y},u_{z},B_{y},B_{z},p)_{t=0}=
\begin{cases}
(1,0,0,0,1,0,1), & -0.5<x<0.5 \\
(0.125,0,0,0,-1,0,1), &otherwise.
\end{cases} \\
\end{flalign*}
\noindent
Where $B_{x}=1$. Under these conditions, we performed calculations using the numerical flux based on Flux NO, and the results are shown in Figure \ref{1D_brio}. As can be seen from the figure, our model is able to approximate compound waves, rarefaction waves, and shock waves to some extent. Although Flux NO was able to capture shock waves, numerical instability issues prevented long-term inference. For the two-dimensional problem, we addressed the Orszag-Tang problem. The initial conditions for the Orszag-Tang problem are described as follows:
\begin{flalign*}
\rho(x,y,0)=\gamma^{2},\quad v_{x}(x,y,0)=-\sin{y},\quad v_{y}(x,y,0)=\sin{x}, \\
p(x,y,0)=\gamma,\quad B_{x}(x,y,0)=-\sin{y}, \quad B_{y}(x,y,0)=\sin{2x}, u_{z}(x,y,0)=B_{z}(x,y,0)=0.
\end{flalign*}
\noindent
When given the initial conditions of the Orszag-Tang problem, the solution at $t=0.25$ was used as the initial condition for the Flux NO scheme, and under these initial conditions, iterations were run 50 ($t=0.5$), 100 ($t=0.75$), 350 ($t=2.0$), and 550 ($t=3.0$) times, with results shown in Figures \ref{2D_short_wzt} and \ref{2D_long_wzt}. Similar to the results in section 4.2, the model approximates well for short-term inference below the time length of the training sample, but accuracy decreases for long-term inference, though the trend is generally followed. It was also observed that the error magnitude increases near shock waves. The values of $\rho$ at the cross-section of $y=0.625\pi$ and how they evolve over time are shown in Figure \ref{2D_section_wzt}. Given that it performs well in short-term prediction, we conducted a further experiment where we used the values at each time point of the Orszag-Tang problem as initial conditions to predict the values after a $\Delta t=0.5$ time. The results of this are shown in Figure \ref{2D_wzt_succ}, and the errors compared to the reference are shown in Figure \ref{2D_wzt_succ_err}. As can be seen from the figures, when predictions are made in the short-term, the errors are significantly reduced. Considering the good generalization performance and short-term approximation ability of the Flux NO model compared to the typical Neural Operator, we can apply the Flux NO model as a surrogate for numerical schemes in short-term inference.

\begin{figure}
  \makebox[\textwidth][c]{\includegraphics[width=0.9\textwidth]{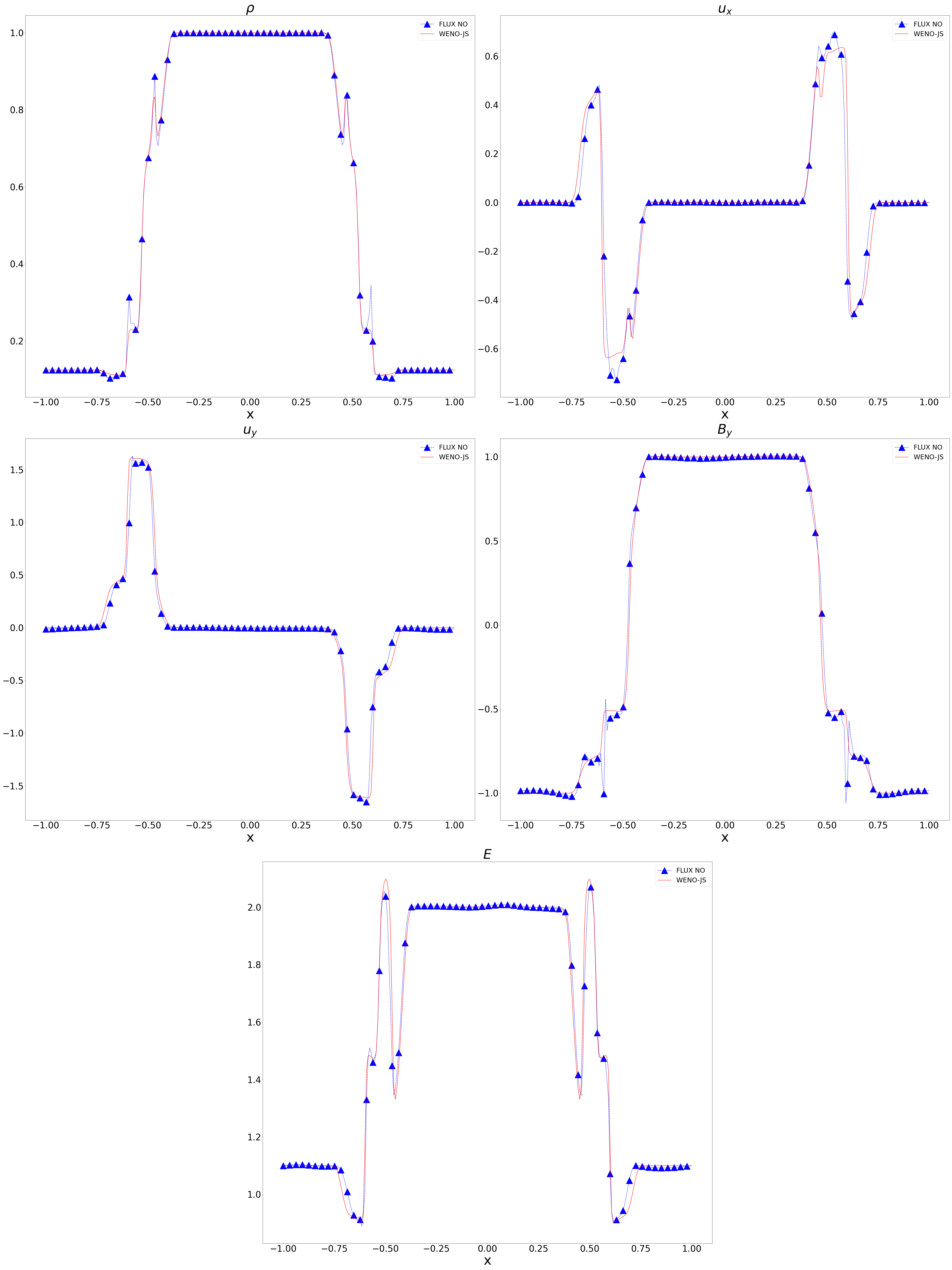}}%
  \caption{Snapshot of the Output from Flux NO and Reference Data at $t=0.007$ for the Modified Brio-Wu Shock Tube Problem.}
  \label{1D_brio}
\end{figure}

\begin{figure}
  \makebox[\textwidth][c]{\includegraphics[width=1.1\textwidth]{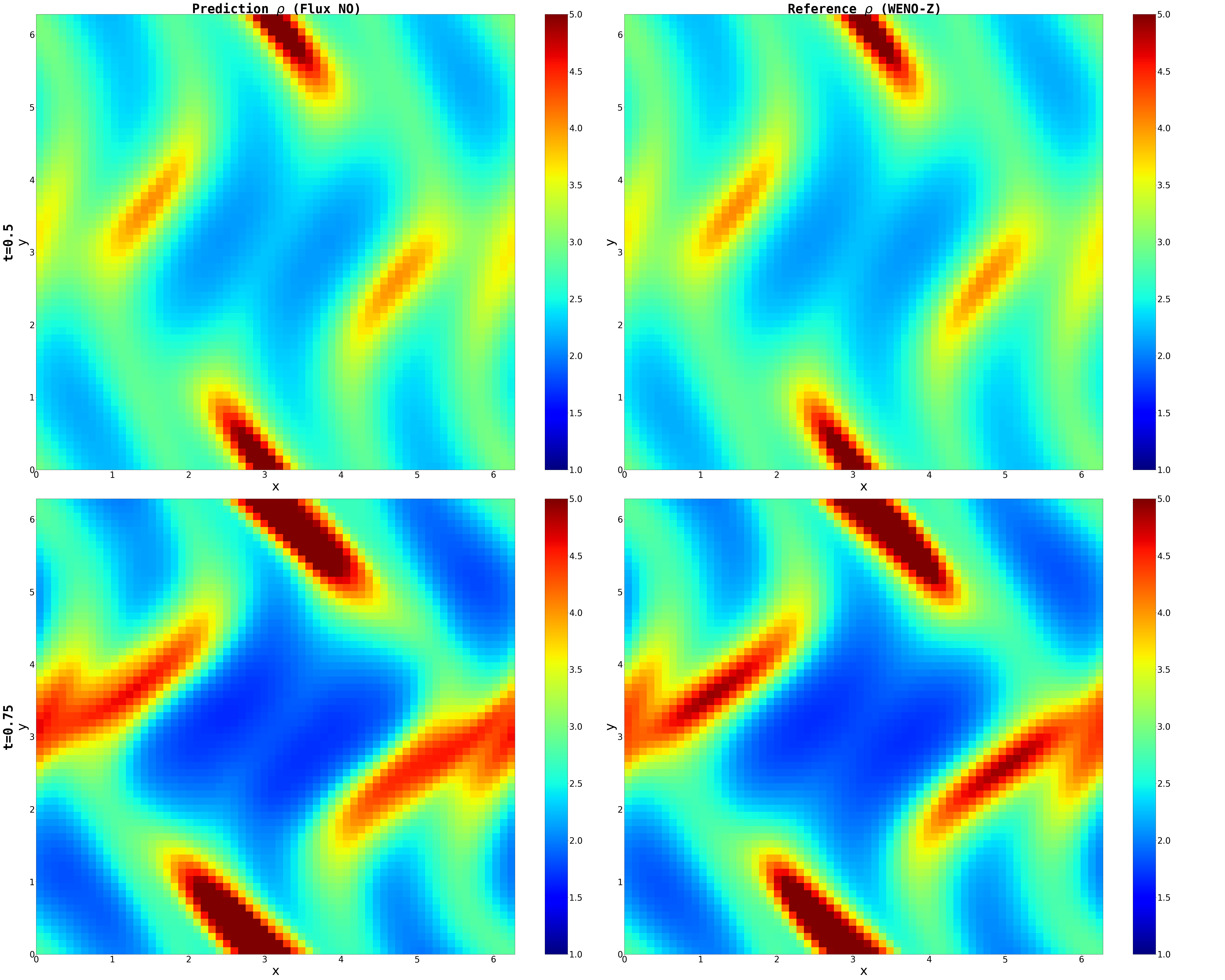}}%
  \caption{Snapshot of the Output from Flux NO and Reference Data at $t=0.5$ and $t=0.75$ for the Orszag-Tang Vortex Problem.}
  \label{2D_short_wzt}
\end{figure}

\begin{figure}
  \makebox[\textwidth][c]{\includegraphics[width=1.0\textwidth]{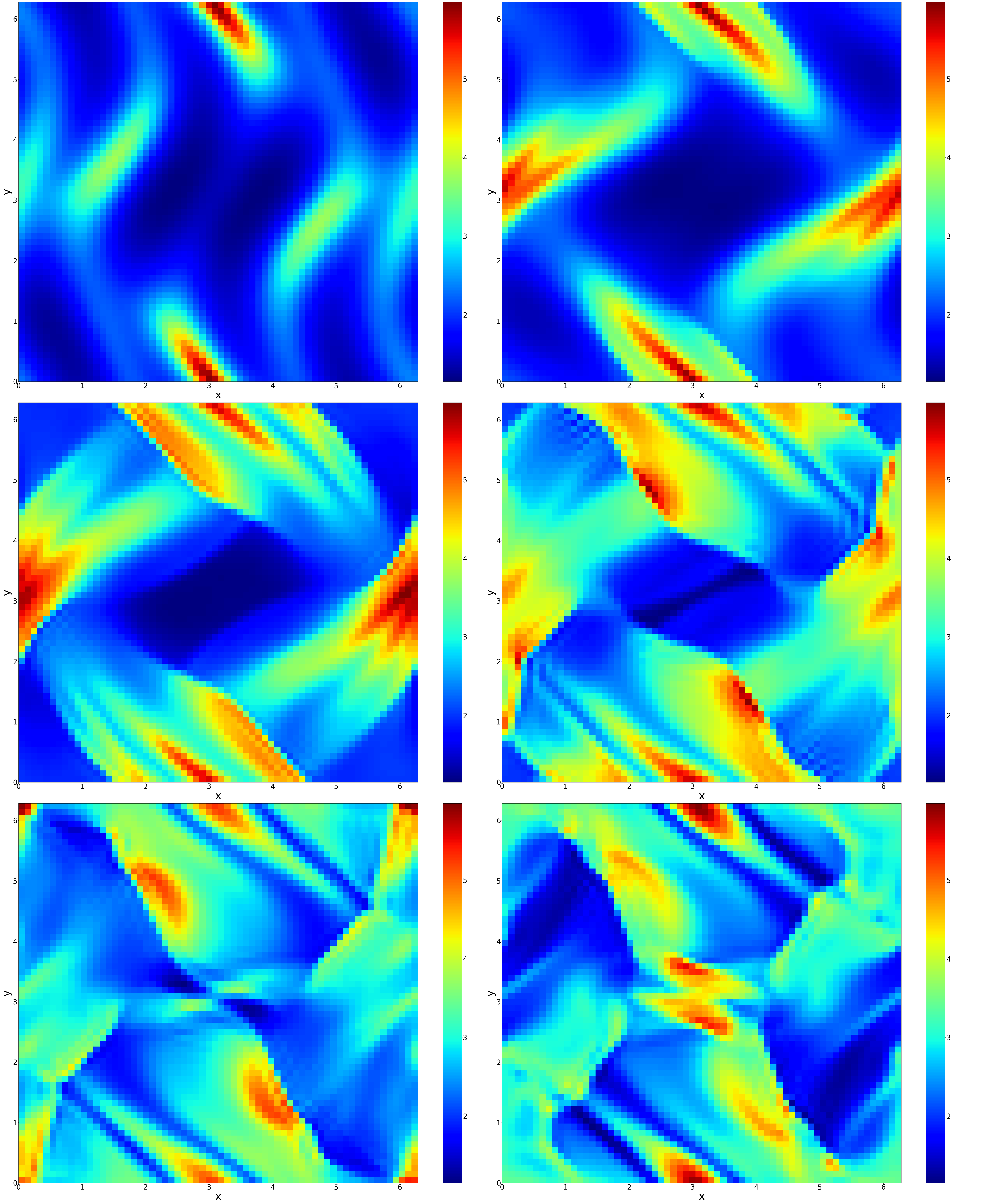}}%
  \caption{Snapshots of the Output from Flux NO and Reference Data at $t=0.5$ (top left), $t=1.0$ (top right), $t=1.5$ (middle left), $t=2.0$ (middle right), $t=2.5$ (bottom left), and $t=3.0$ (bottom right) for the Orszag-Tang Vortex Problem. Each result represents the solution after 50 iterations calculated with Flux NO for the given initial conditions.}
  \label{2D_wzt_succ}
\end{figure}

}
\subsection{Comparison with Other Methods}{

In this section, we demonstrate both quantitatively and qualitatively that the Flux NO model possesses superior generalization abilities compared to traditional FNO models. Like Flux NO, both FNO 2D and FNO 3D use frequencies up to 4 on each axis and have a width of 72. FNO 2D operates by predicting an 8-dimensional vector value after a $\Delta t = 0.5$ given an input function, while FNO 3D works by producing the next 25 snapshots simultaneously from a 4D image of dimensions 25 by 64 by 64 by 8, considering the time axis in the dataset. For FNO 2D, we also trained a larger model, FNO 2D (heavy), with 8 frequencies and a width of 104. The dataset was adapted from the one used for training Flux NO, structured with 10,000 function pairs for 2D FNO and 500 pairs for 3D FNO. All models used the same learning rate and weight decay settings of 1e-3 and 1e-4 respectively, with Adam as the optimizer and StepLR as the scheduler with a step\_size of 200 and gamma of 0.5. Training was conducted over 1000 epochs, with a batch size of 500 for 2D FNO and 25 for 3D FNO. Experimental results on the test dataset are presented in Table \ref{t4}, showing Flux NO's significantly superior performance, with 3D FNO being the better performer among the remaining models. To quantitatively and qualitatively compare 3D FNO and Flux NO, we tasked 3D FNO with inferring solutions for the Orszag-Tang problem. The results for Flux NO and 3D FNO are detailed in Table \ref{t5}, with qualitative illustrations in Figures \ref{2D_short_wzt}, \ref{2D_long_wzt}, and \ref{2D_short_wzt_3D}. As the table and figures illustrate, unlike Flux NO, 3D FNO fails to perform effectively on out-of-distribution samples and in long-term inference scenarios.

\begin{table}[H]
\makebox[\textwidth][c]{
\begin{tabular}{lllll}
\hline\noalign{\smallskip}
\thead{(M, SD)} & \thead{t=0.5} & \thead{t=0.75} & \thead{t=1.75} & \thead{t=2.0} \\
\hline
\thead{rel $l^{2}$ Flux NO} & (\textbf{2.5e-3}, 1.11e-1)  & (\textbf{6.6e-3}, 1.11e-1) & (\textbf{2.63e-2}, 1.11e-1) & (\textbf{3.37e-2}, 1.11e-1) \\
\hline
\thead{rel $l^{\infty}$ Flux NO} & (\textbf{1.5e-3}, 1.10e-1)  & (\textbf{2.5e-3}, 1.11e-1) & (\textbf{4.27e-2}, 1.11e-1) & (\textbf{4.4e-2}, 1.11e-1) \\
\hline
\thead{rel $l^{2}$ 2D FNO} & (2.39e-1, 2.91e-2) & (8.02e-1, 3.64e-2) & (8.14e-1, 5.54e-2) & (7.70e-1, 1.99e-2)\\ 
\hline
\thead{rel $l^{\infty}$ 2D FNO} & (1.95e-1, 6.01e-2) & (7.71e-1, 1.30e-1) & (8.83e-1, 3.08e-1) & (6.91e-1, 1.25e-1)\\ 
\hline
\thead{rel $l^{2}$ 2D FNO (heavy)} & (1.51e-1, 1.71e-2) & (7.44e-1, 4.29e-2) & (7.53e-1, 2.9e-2) & (7.42e-1, 2.66e-2)\\ 
\hline
\thead{rel $l^{\infty}$ 2D FNO (heavy)} & (6.59e-2, 3.59e-2) & (6.97e-1, 1.03e-1) & (6.73e-1, 1.06e-1) & (6.67e-1, 1.29e-1)\\ 
\hline
\thead{rel $l^{2}$ 3D FNO}  & (1.42e-1, 1.46e-2)  & (3.09e-1, 3.68e-2) & (6.42e-1, 6.29e-2) & (6.83e-1, 5.46e-2)\\ 
\hline
\thead{rel $l^{\infty}$ 3D FNO}  & (8.22e-2, 7.41e-2)  & (2.26e-1, 1.44e-1) & (1.02, 8.74e-1) & (1.07, 6.21e-1)\\ 
\hline

\end{tabular}
}
\caption{Comparisons of Flux NO with Other Standard FNO Models: Means (M) and Standard Deviations (SD) of Relative $l^{2}$ and $l^{\infty}$ Norms Between the Outputs from Each Model and Reference Data at Various Times in the Two-Dimensional Case, Across 10 Test Samples.} \label{t4}

\end{table}

\begin{table}[H]
\centering
\begin{tabular}{lllll}
\hline\noalign{\smallskip}
& \thead{t=0.5} & \thead{t=0.75} & \thead{t=1.75} & \thead{t=2.0} \\
\hline
\thead{rel $l^{2}$ Flux NO} & \bf{4.42e-2}  & \bf{8.38e-2} & \bf{2.28e-1} & \bf{2.53e-1} \\
\hline
\thead{rel $l^{\infty}$ Flux NO} & \bf{1.11e-2}  & \bf{2.08e-2} & \bf{3.13e-1} & \bf{5.39e-1} \\
\hline
\thead{rel $l^{2}$ 3D FNO}  & 1.36e-1  & 3.73e-1 & 1.03 & 1.06 \\ 
\hline
\thead{rel $l^{\infty}$ 3D FNO}  & 5.59e-2  & 4.04e-1 & 7.72 & 10.32 \\ 
\hline

\end{tabular}
\caption{Comparisons of Flux NO with 3D FNO: Relative $l^{2}$ and $l^{\infty}$ Norms Between the Outputs from Each Model and Reference Data at Various Times in the Two-Dimensional Case, on the Orszag-Tang Problem.} \label{t5}
\end{table}

}

\subsection{Memory and Time costs}{

In this section, we analyze the memory requirements, specifically the number of parameters, and the inference time required for each model. Experimental results for each model are presented in Table \ref{t6}, where WENO-Z calculations were performed on a CPU with Fortran language, while the rest of the neural operator models were computed on a GPU basis. As indicated in the table, assuming efficient GPU computation parallelization, our Flux NO is approximately 25 times faster than the WENO-Z method. Furthermore, when the WENO-Z method was run under the same GPU conditions (using PyTorch) as Flux NO, it was observed to be 250 times slower than Flux NO. From this, we can predict that Flux NO, when optimized to operate in environments like WENO-Z, would allow for significantly faster inference. The inference times in this section were measured under the conditions used for generating samples in Section 4.2. Even with higher resolution images, the computational complexity of Flux NO is inferred to depend simply on the lifting, projection layer's FCN and Fourier layer's CNN layers, and the computation of the Fast Fourier Transform due to the limited frequency usage in the Fourier layer. Because Flux NO inherently performs calculations locally, it naturally takes more time to reach the same point compared to the traditional FNO. However, unlike FNO, which can only infer fixed distributions at fixed time intervals, Flux NO offers the advantage of flexible time interval selection, allowing for continuous inference. As shown in Table \ref{t6} and the results in Section 4.4, despite having a similar number of parameters, 2D FNO (heavy) and 3D FNO show inferior generalization capabilities compared to Flux NO. Considering these factors, Flux NO can be seen as embodying the strengths of both numerical schemes—generalization capability and robustness—and the fast computational abilities of neural operators.

\begin{table}[H]
\centering
\begin{tabular}{lll}
\hline\noalign{\smallskip}
 \thead{Models}   & \thead{Inference Time \\ for $\Delta t = 0.5$ } & \thead{Number of Parameters} \\
\hline
\thead{Flux NO}  & 4.16e-1s (4.16e-3s) & 8,204,176 \\
\hline
\thead{2D FNO}  & 3.04e-3s & 1,022,264 \\
\hline
\thead{2D FNO (heavy)} & 4.74e-3s & 8,355,064 \\
\hline
\thead{3D FNO}  & 1.81e-2s (9.05e-3s) & 7,990,064 \\ 
\hline
\thead{WENO-Z}   & 1.05e+1s (1.05e-1s) &   \\ 
\hline
\thead{WENO-Z (on PyTorch)}   & 9.98e+1s (9.98e-1s) &   \\ 
\hline

\end{tabular}
\caption{Comparisons of Flux NO with Other Standard FNO Models: Memory Requirements and Inference Times. The time in parentheses represents the inference time for a single run.} \label{t6}
\end{table}

}

\subsection{Ablation Study}{

In this section, we conduct an ablation analysis to explore the effects of the additional loss functions we designed. Models were trained with both TVD (Total Variation Diminishing) loss and divergence-free loss selectively removed, and the impact on $\nabla \cdot \bold{B}$ and Total Variation was observed over 75 iterations across 10 test samples. The results are presented in Figure \ref{Albations}, which illustrates the influence of each loss on the corresponding values of $\nabla \cdot \bold{B}$ and Total Variation. Relaxing $\nabla \cdot \bold{B}$ is a significant factor in obtaining physical solutions and also affects the stability when computing time steps adaptively. Mitigating the increase in Total Variation is likewise related to the stability of the numerical scheme. We also evaluated the performance of our designed loss function and model architecture on test samples by conducting ablation studies. The non-adapted model, referred to as Flux NO (non-splitted), handles all physical variables simultaneously instead of assigning separate neural operator components to each variable. The neural operator components of Flux NO (non-splitted) consist of two models, each corresponding to flux along an axis. This model increases the frequency usage to 6 and width to 96 to enhance expressiveness, unlike the adapted Flux NO. Training datasets and hyperparameters remain consistent with the baseline. Results of the ablation are presented in Table \ref{t7}. As the table illustrates, generally removing our designed loss and modifications in the model architecture results in decreased performance for long-term inference. In the case of the $l^{\infty}$ loss ablated model, although the ablated model shows better performance in the $l^{2}$ norm, it exhibit inferior performance in the $l^{\infty}$ norm for long-term inference compared to the baseline model.

\begin{figure}[H]

     \begin{center}

        \includegraphics[width=0.45\textwidth]{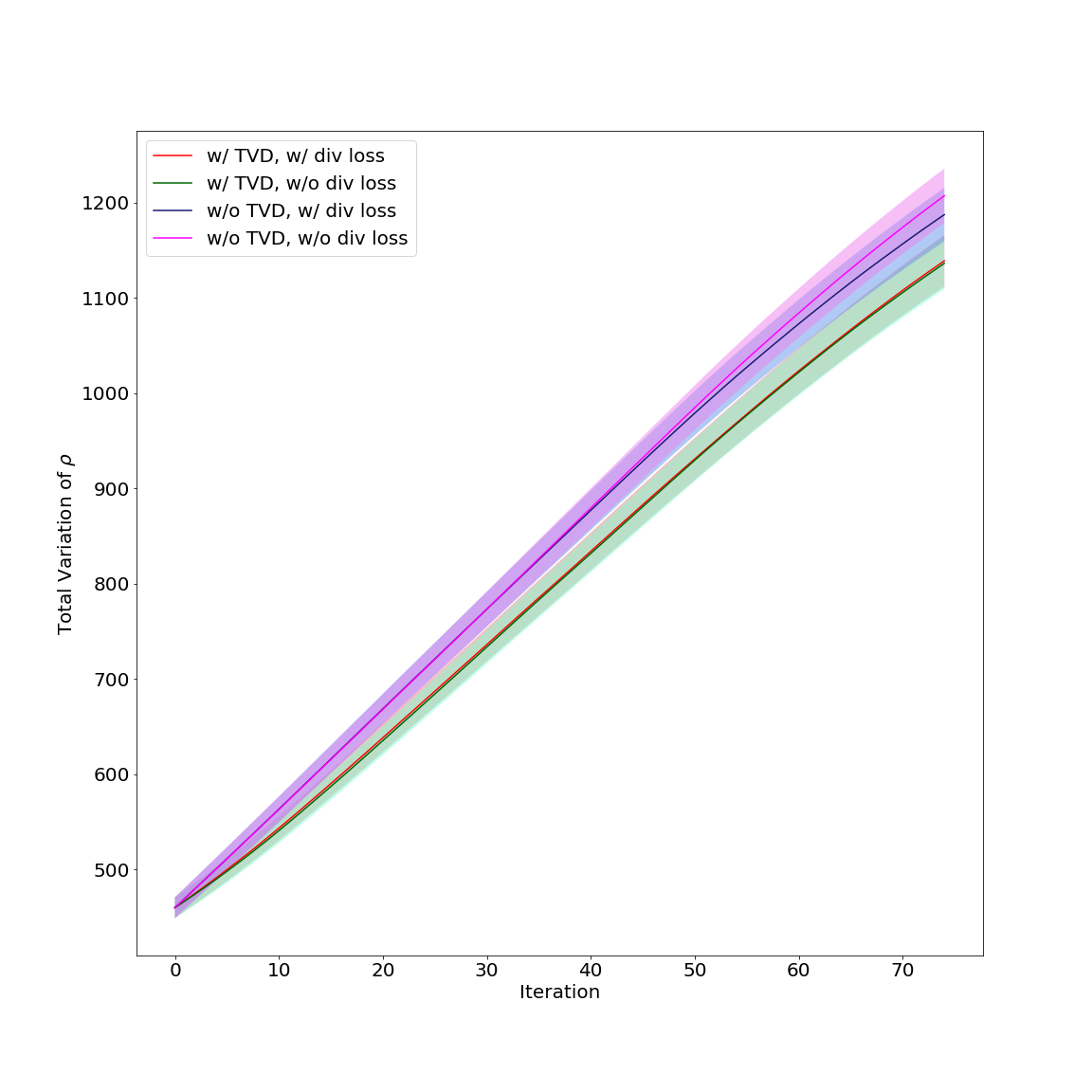}
        \includegraphics[width=0.45\textwidth]{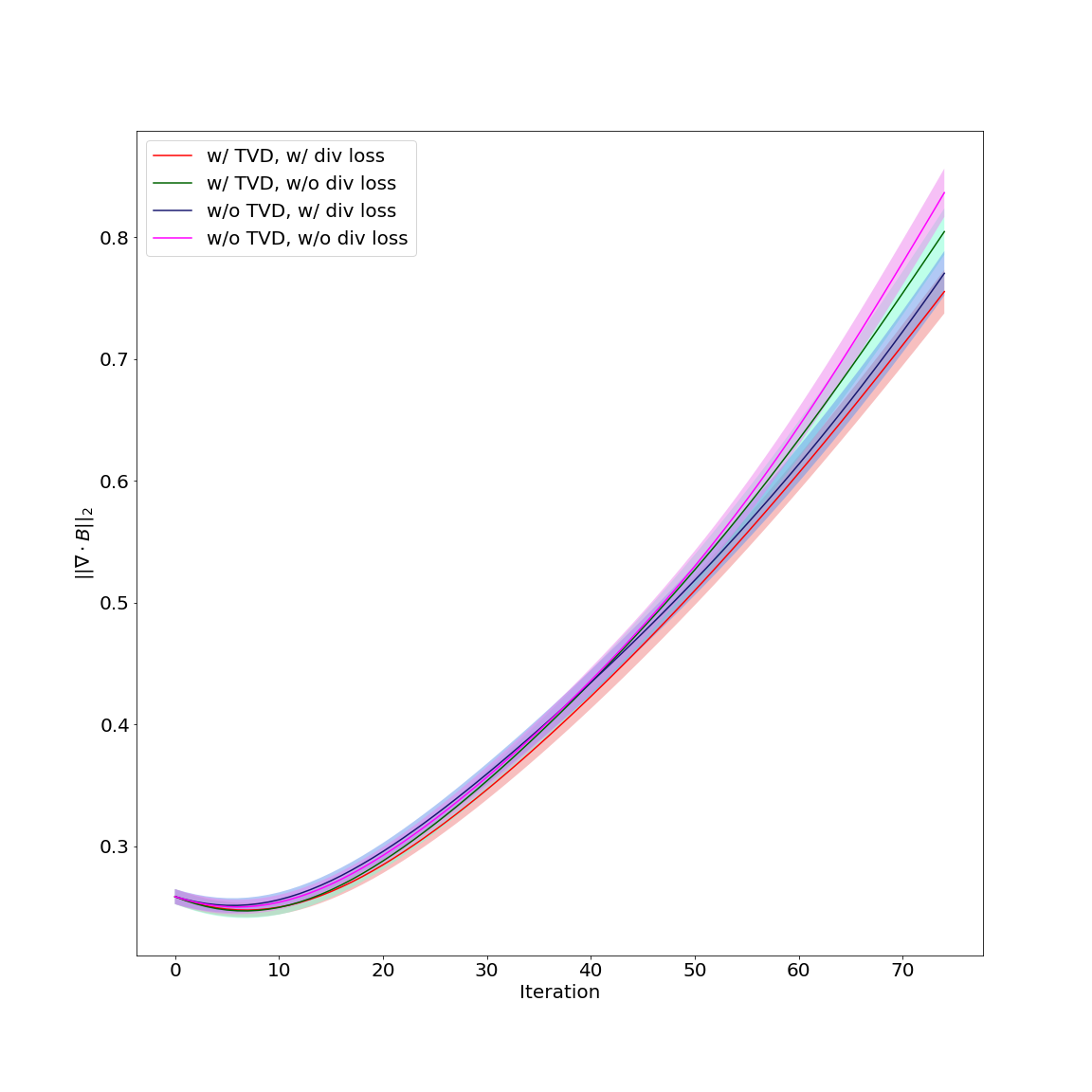}
    
   \end{center}
    \caption{%
      Graphical Representation of Means (M) and Standard Deviations (SD) of Total Variation (left) and $\|\nabla\cdot \bold{B}\|_{2}$ (right) Values Across 10 Test Samples Along the Time-Marching Iterations. The shaded area represents the scaled standard deviation, which is divided by 100, and the solid line represents the mean.}
   \label{Albations}
\end{figure}

\begin{table}[H]
\makebox[\textwidth][c]{
\begin{tabular}{lllll}
\hline\noalign{\smallskip}
\thead{(M, SD)} & \thead{t=0.5} & \thead{t=0.75} & \thead{t=1.75} & \thead{t=2.0} \\
\hline
\thead{rel $l^{2}$ Flux NO} & (8e-4, 2.04e-2)  & (1.4e-3, 2.04e-2) & (3.9e-3, 2.04e-2) & (4.9e-3, 2.04e-2) \\
\hline
\thead{rel $l^{\infty}$ Flux NO} & (2e-4, 2.04e-2)  & (4e-4, 2.04e-2) & (3.8e-3, 2.04e-2) & (3.9e-3, 2.04e-2) \\
\hline
\thead{rel $l^{2}$ Flux NO (non-splitted)} & (8e-4, 2.04e-2) & (1.6e-3, 2.04e-2) & (5.6e-3, 2.04e-2) & (7e-3, 2.04e-2)\\ 
\hline
\thead{rel $l^{\infty}$ Flux NO (non-splitted)} & (2e-4, 2.04-2) & (4e-4, 2.04e-2) & (7e-3, 2.04e-2) & (1.13e-2, 2.04e-2)\\ 
\hline
\thead{rel $l^{2}$ Flux NO w/o $l^{\infty}$ norm} & (8e-4, 2.04e-2) & (1.3e-3, 2.04e-2) & (3.8e-3, 2.04e-2) & (4.5e-3, 2.04e-2)\\ 
\hline
\thead{rel $l^{\infty}$ Flux NO w/o $l^{\infty}$ norm} & (1e-4, 2.04e-2) & (3e-4, 2.04e-2) & (5.1e-3, 2.04e-2) & (5.1e-3, 2.04e-2)\\ 
\hline
\thead{rel $l^{2}$ Flux NO w/o TVD} & (8e-4, 2.04e-2) & (1.3e-3, 2.04e-2) & (4.4e-3, 2.04e-2) & (5.2e-3, 2.04e-2)\\ 
\hline
\thead{rel $l^{\infty}$ Flux NO w/o TVD} & (2e-4, 2.04e-2) & (1e-3, 2.04e-2) & (1.26e-2, 2.04e-2) & (5e-3, 2.04e-2)\\ 
\hline
\thead{rel $l^{2}$ Flux NO w/o div loss} & (9e-4, 2.04e-2) & (1.6e-3, 2.04e-2) & (4.1e-3, 2.04e-2) & (4.9e-3, 2.04e-2)\\ 
\hline
\thead{rel $l^{\infty}$ Flux NO w/o div loss} & (2e-4, 2.04e-2) & (5e-4, 2.04e-2) & (3e-3, 2.04e-2) & (4.3e-3, 2.04e-2)\\ 
\hline

%\thead{rel $l^{2}$ Flux NO w/o TVD, w/o div loss} & (8e-4, 2.04e-2) & (1.4e-3, 2.04e-2) & (4.1e-3, 2.04e-2) & (4.8e-3, 2.04e-2)\\ 
%\hline
%\thead{rel $l^{\infty}$ Flux NO w/o TVD, w/o div loss} & (1e-4, 2.04e-2) & (5e-4, 2.04e-2) & (2.7e-3, 2.04e-2) & (2.9e-3, 2.04e-2)\\ 
%\hline

\end{tabular}
}
\caption{Comparisons of Flux NO with Ablated Models Across 50 Test Samples.} \label{t7}

\end{table}

}

\section{Conclusion}
In this study, we propose a method to solve the one-dimensional and two-dimensional cases of the ideal MHD, one of the equations describing plasma, using our newly designed Flux NO technique. We have demonstrated through experiments that Flux NO can be applied not only to the one-dimensional scalar conservation law (\cite{Kim:24}) but also to the more complex ideal MHD equations, exhibiting long-term inference capabilities and generalization ability on out-of-distribution samples. We implemented specific loss functions to satisfy the TVD properties and divergence-freeness of the solutions, and adapted the model architecture to allocate different models to handle various variables appropriately. Our designed Flux NO shows flexibility in these modifications, particularly in the potential for adapting or improving Neural Operator component in the scheme to suit the problem. Although Flux NO requires more iterations to achieve the same solution, resulting in longer inference times compared to other Neural Operators, it allows for continuous inference and notably shorter inference times than classical numerical schemes. The dataset we adopted in this research is quite limited, and computing resources were also constrained; nonetheless, it is noteworthy that we achieved such generalization performance. Flux NO has the potential for further advancement and broadening of its application scope with more diverse datasets, larger-scale computing resources, and modifications to the loss functions and architecture.

% Acknowledgements should go at the end, before appendices and references

%\acks{ }


\newpage
\appendix
%%\section*{Appendix A.}
\section{Error Plots}

\begin{figure}[H]
  \makebox[\textwidth][c]{\includegraphics[width=1.3\textwidth]{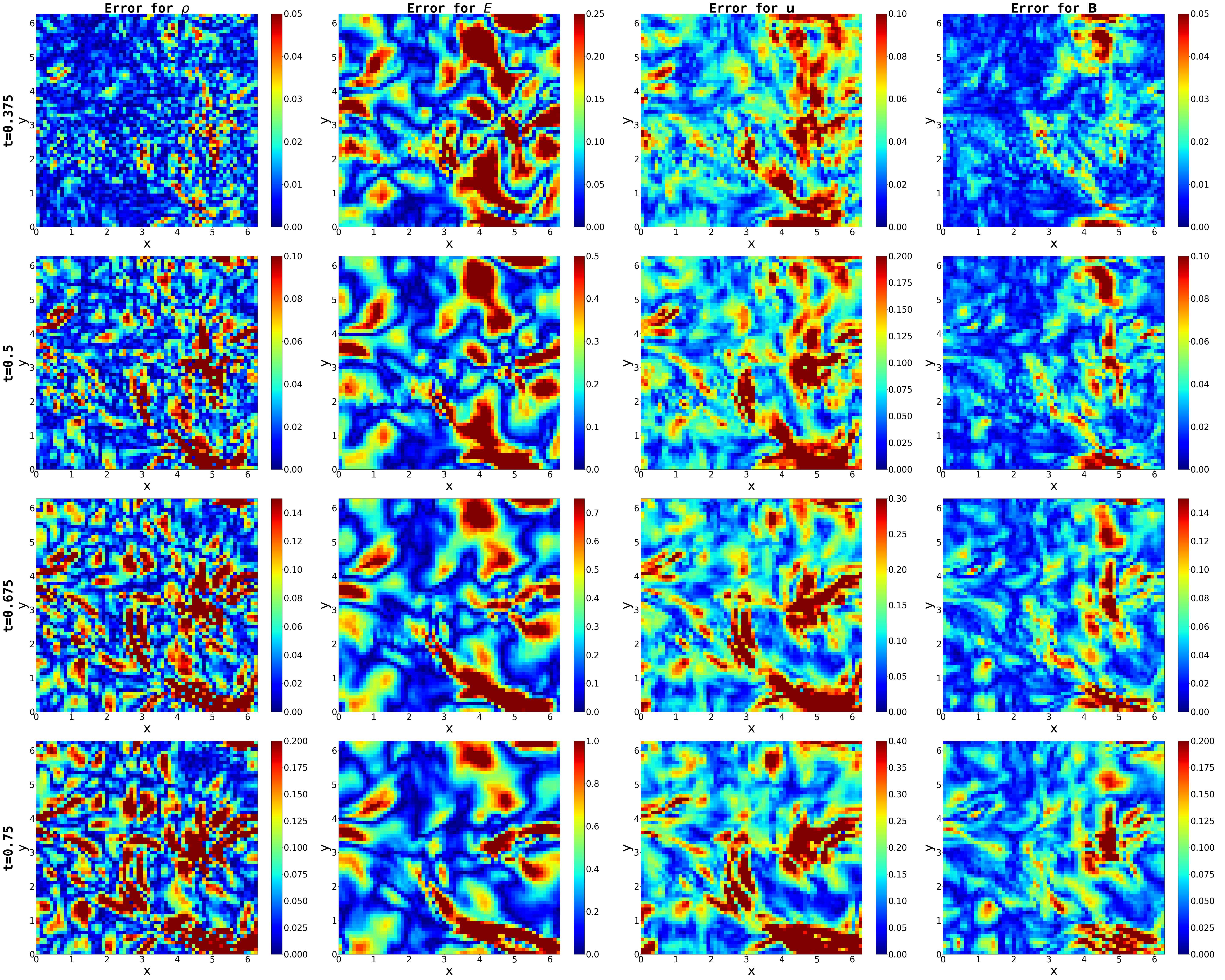}}%
  \caption{Short-Term Time Evolution of Relative Errors for Each Component in the Output from Flux NO Compared to Reference Data for Two-Dimensional Ideal MHD.}
  \label{2D_short_Error}
\end{figure}

\begin{figure}[H]
  \makebox[\textwidth][c]{\includegraphics[width=1.3\textwidth]{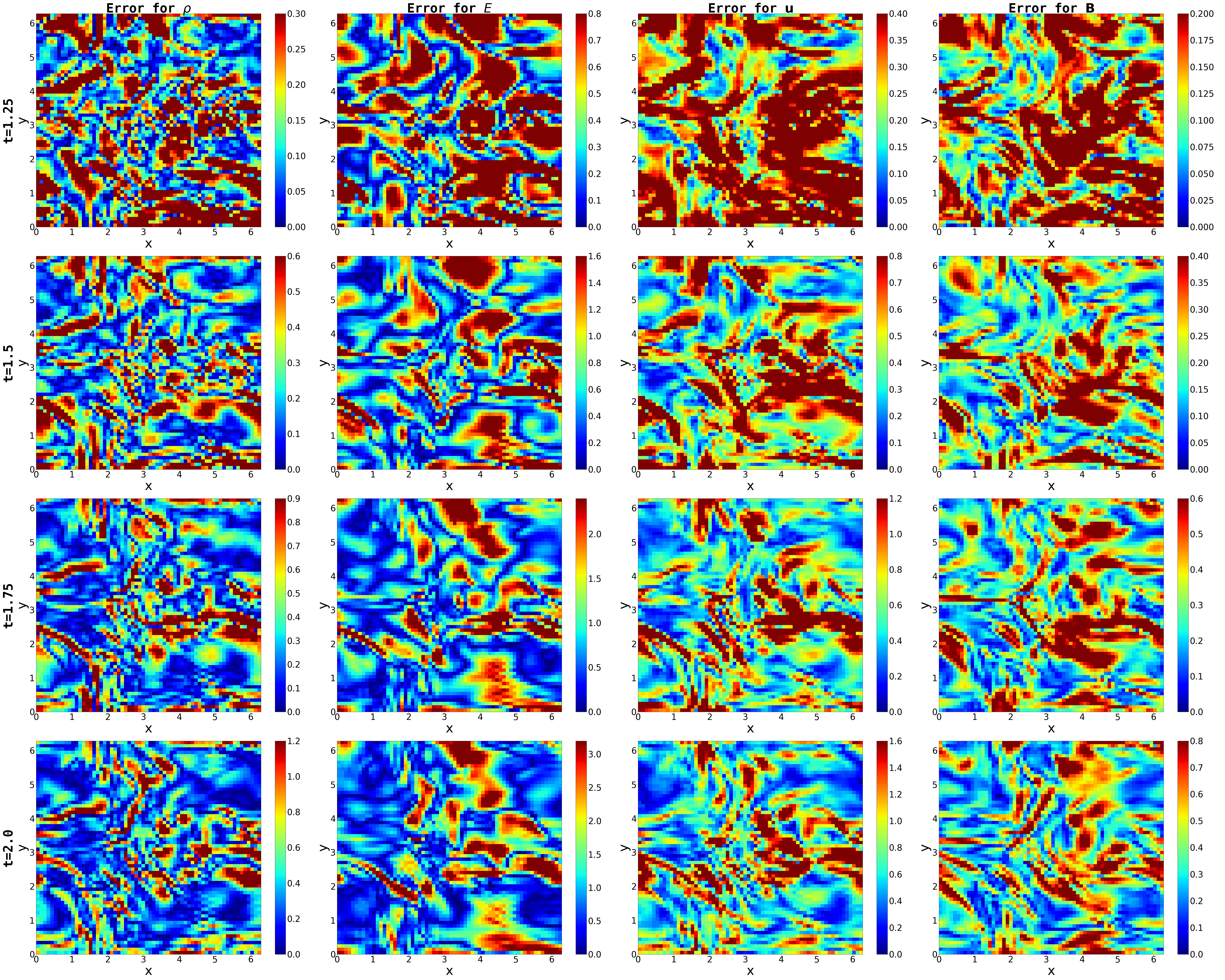}}%
  \caption{Long-Term Time Evolution of Relative Errors for Each Component in the Output from Flux NO Compared to Reference Data for Two-Dimensional Ideal MHD.}
  \label{2D_long_Error}
\end{figure}

\begin{figure}[H]
  \makebox[\textwidth][c]{\includegraphics[width=1.3\textwidth]{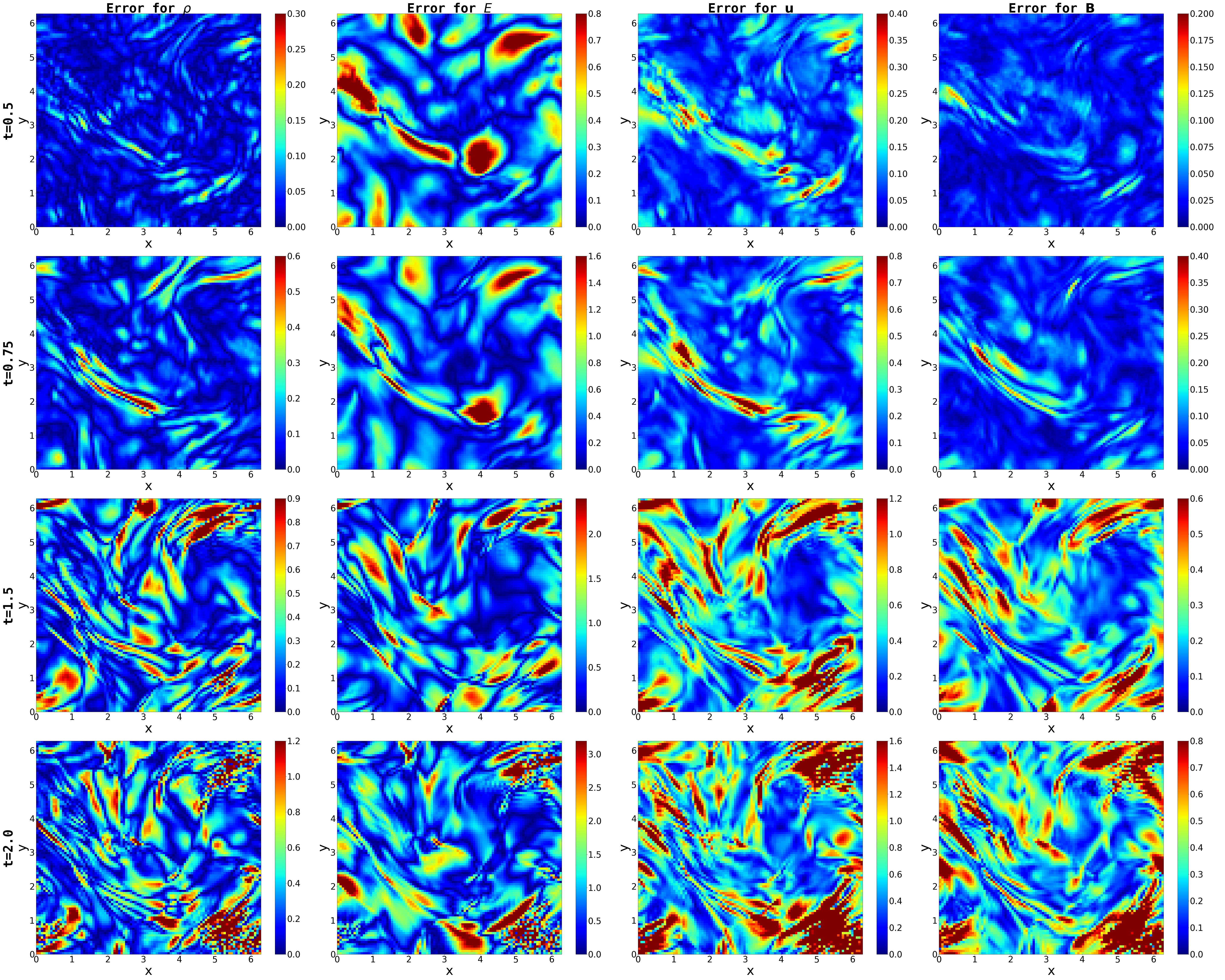}}%
  \caption{Time Evolution of Relative Errors for $\rho$ in the Output from Flux NO Compared to Reference Data in the Higher Resolution (96 by 96) for Two-Dimensional Ideal MHD.}
  \label{2D_high_Error}
\end{figure}

\begin{figure}[H]
  \makebox[\textwidth][c]{\includegraphics[width=1.0\textwidth]{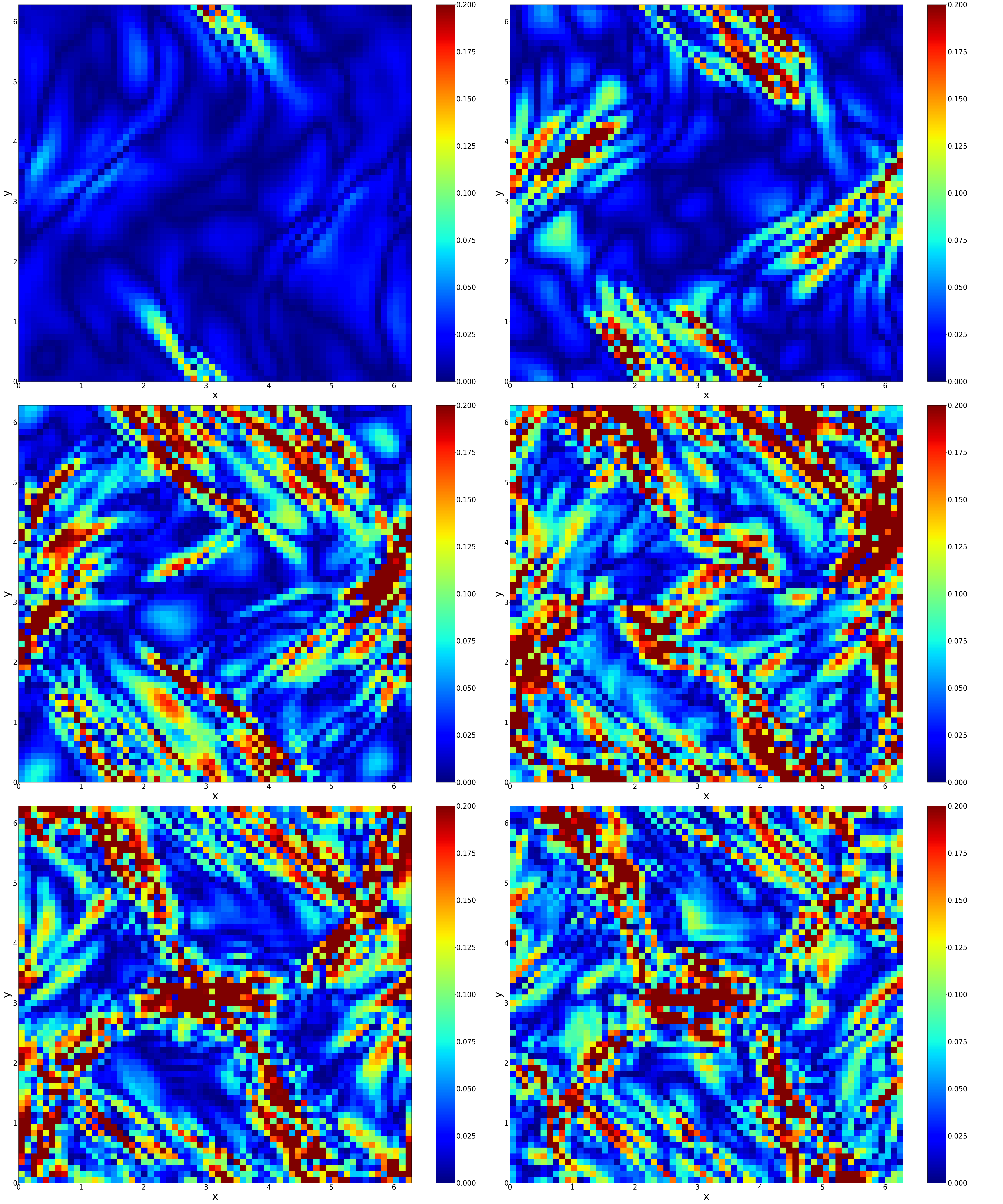}}%
  \caption{Snapshots of the Errors between Output from Flux NO and Reference Data at $t=0.5$ (top left), $t=1.0$ (top right), $t=1.5$ (middle left), $t=2.0$ (middle right), $t=2.5$ (bottom left), and $t=3.0$ (bottom right) for the Orszag-Tang Vortex Problem. Each result represents the solution after 50 iterations calculated with Flux NO for the given initial conditions.}
  \label{2D_wzt_succ_err}
\end{figure}

\section{Supplementary Figures}

\begin{figure}[H]
  \makebox[\textwidth][c]{\includegraphics[width=1.3\textwidth]{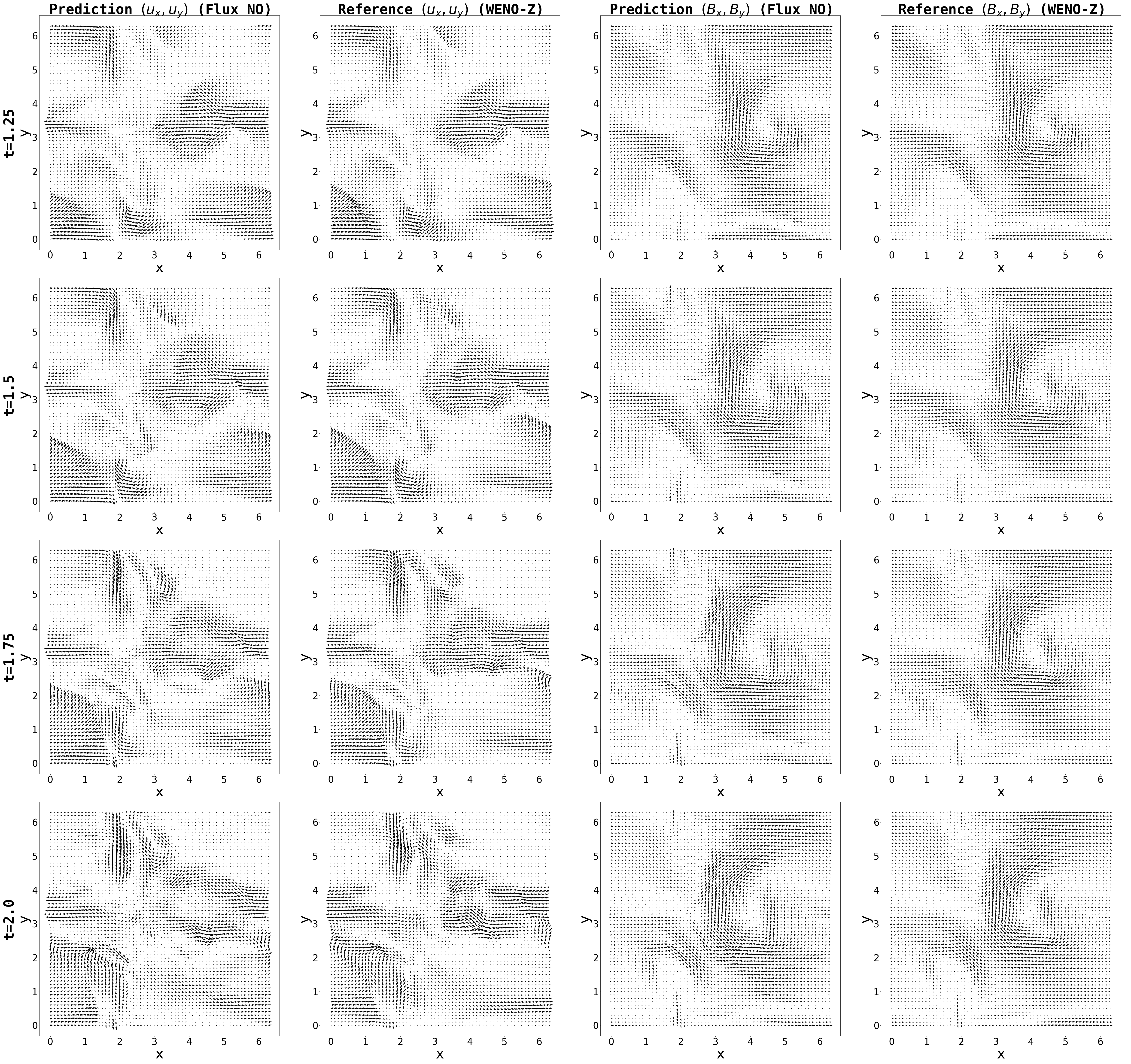}}%
  \caption{Long-Term Time Evolution of $(u_{x},u_{y})$ and $(B_{x},B_{y})$ in the Output from Flux NO Compared to Reference Data for Two-Dimensional Ideal MHD.}
  \label{2D_long_u_B}
\end{figure}

\begin{figure}[H]
  \makebox[\textwidth][c]{\includegraphics[width=1.3\textwidth]{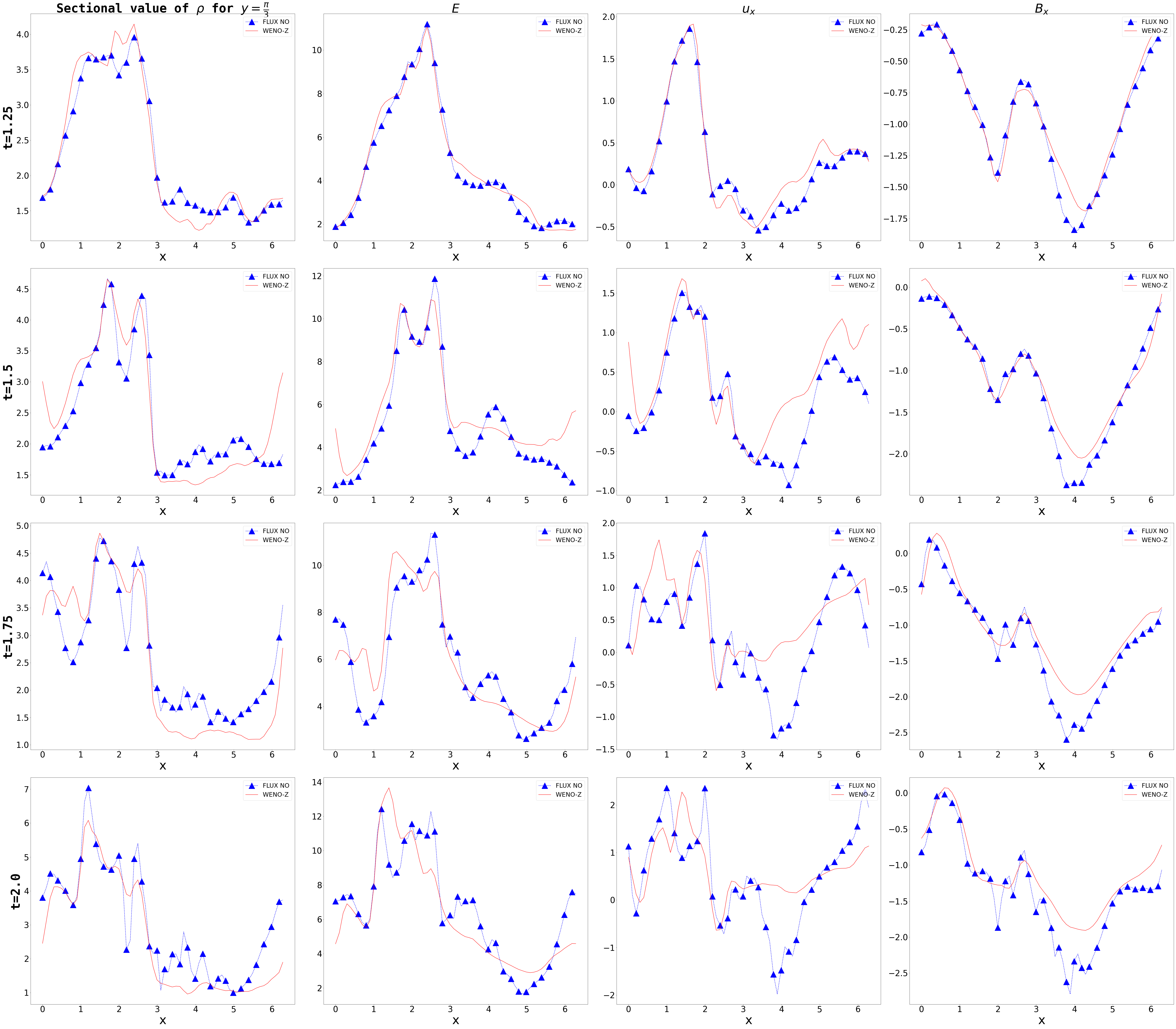}}%
  \caption{Long-Term Time Evolution of Each Component for Section $y=\frac{\pi}{3}$ in the Output from Flux NO Compared to Reference Data for Two-Dimensional Ideal MHD.}
  \label{2D_long_section}
\end{figure}

\begin{figure}[H]
  \makebox[\textwidth][c]{\includegraphics[width=1.3\textwidth]{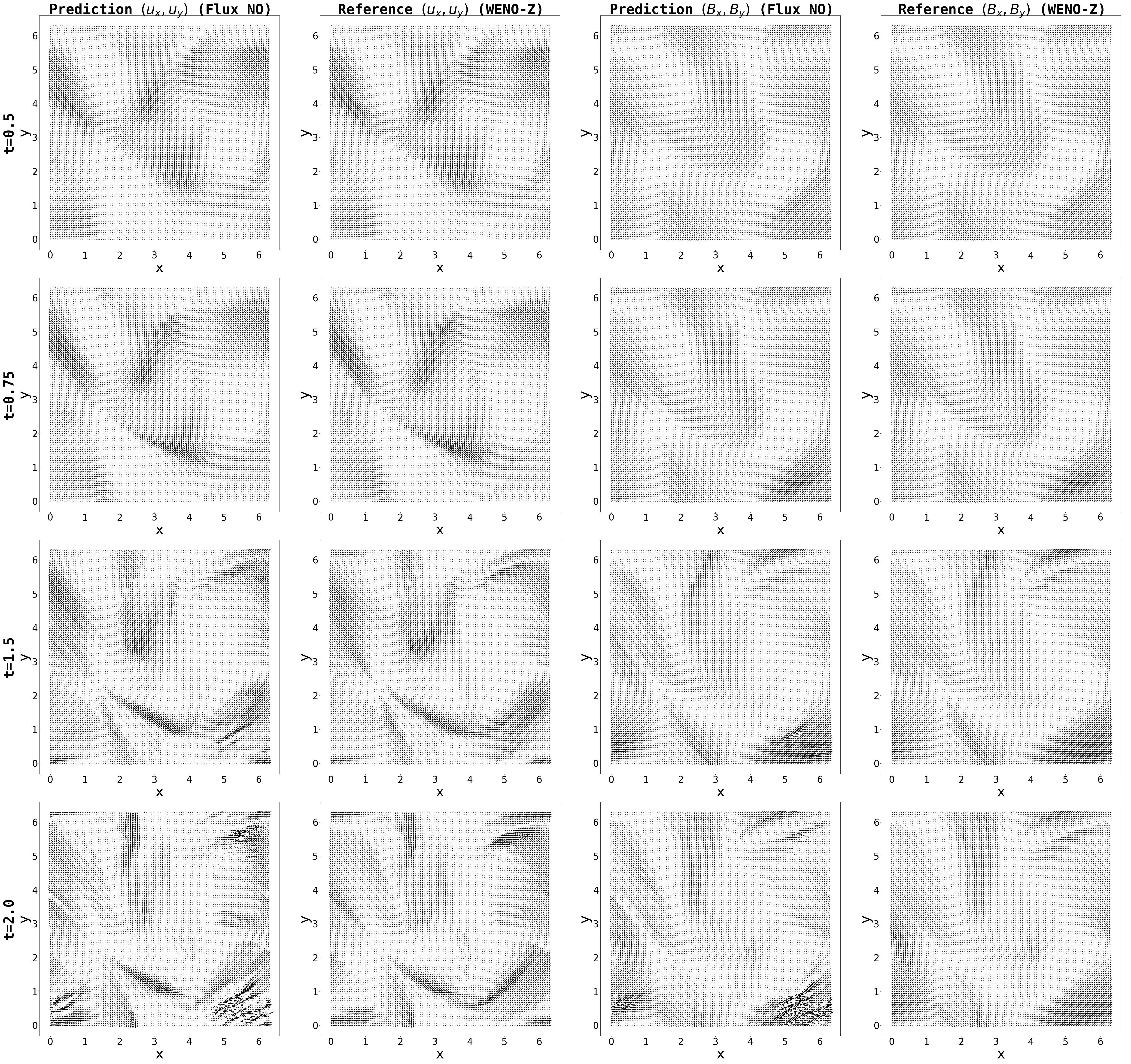}}%
  \caption{Time Evolution of $(u_{x},u_{y})$ and $(B_{x},B_{y})$ in the Output from Flux NO Compared to Reference Data in the Higher Resolution (96 by 96) for Two-Dimensional Ideal MHD.}
  \label{2D_high_u_B}
\end{figure}

\begin{figure}[H]
  \makebox[\textwidth][c]{\includegraphics[width=1.3\textwidth]{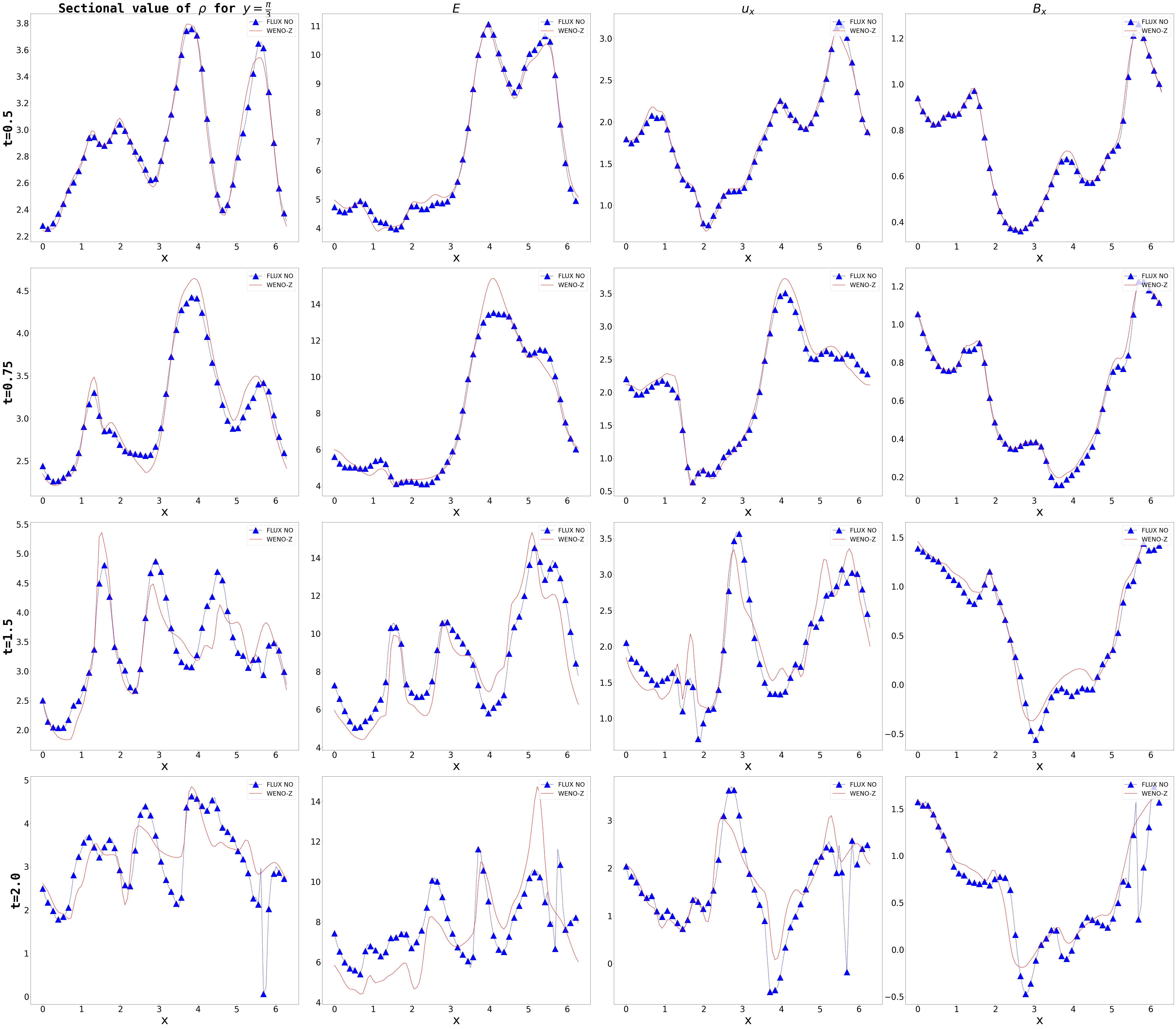}}%
  \caption{Time Evolution of Each Component for Section $y=\frac{\pi}{3}$ in the Output from Flux NO Compared to Reference Data in the Higher Resolution (96 by 96) for Two-Dimensional Ideal MHD.}
  \label{2D_high_section}
\end{figure}

\begin{figure}[H]
  \makebox[\textwidth][c]{\includegraphics[width=1.1\textwidth]{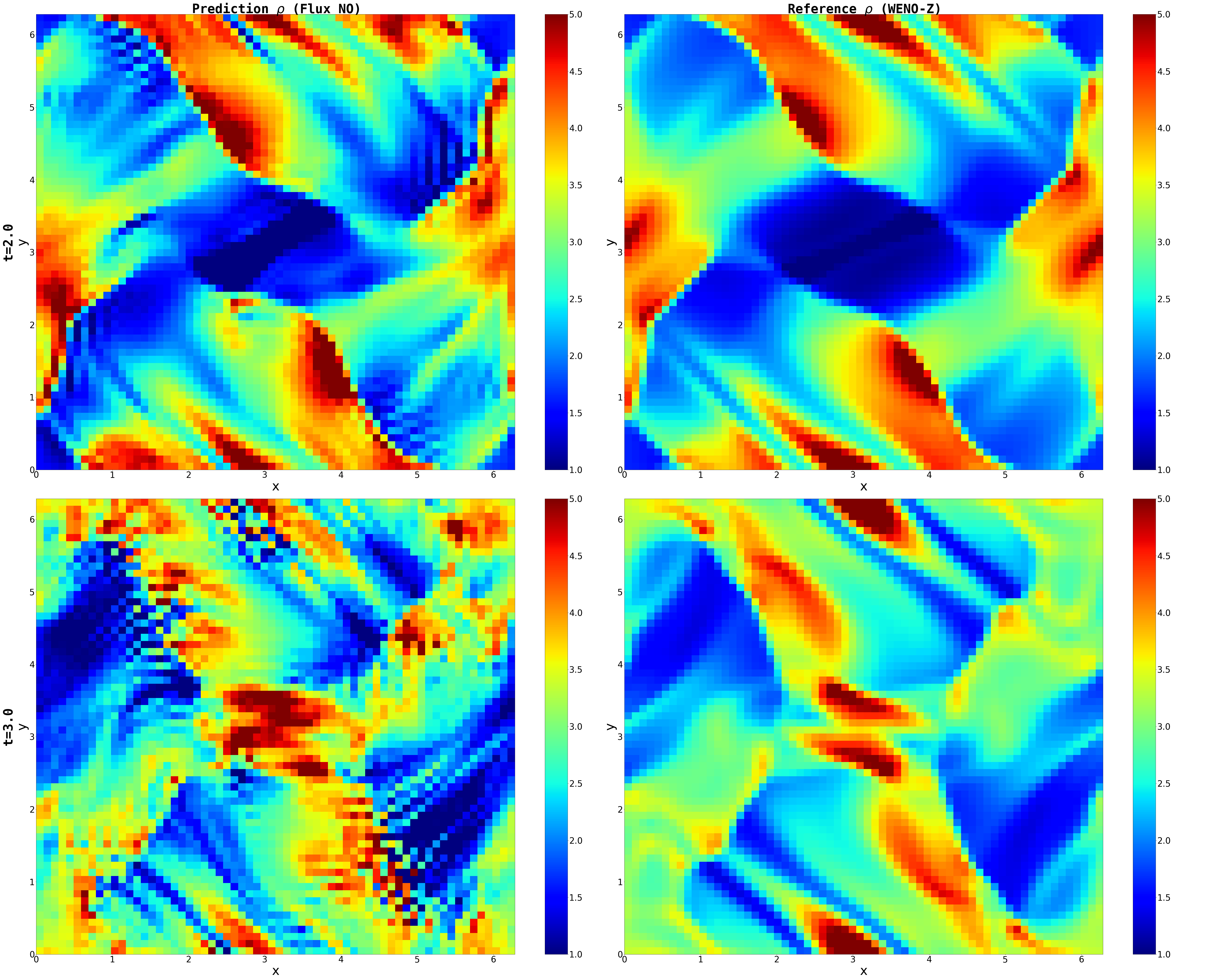}}%
  \caption{Snapshot of the Output from Flux NO and Reference Data at $t=2.0$ and $t=3.0$ for the Orszag-Tang Vortex Problem.}
  \label{2D_long_wzt}
\end{figure}

\begin{figure}[H]
  \makebox[\textwidth][c]{\includegraphics[width=1.1\textwidth]{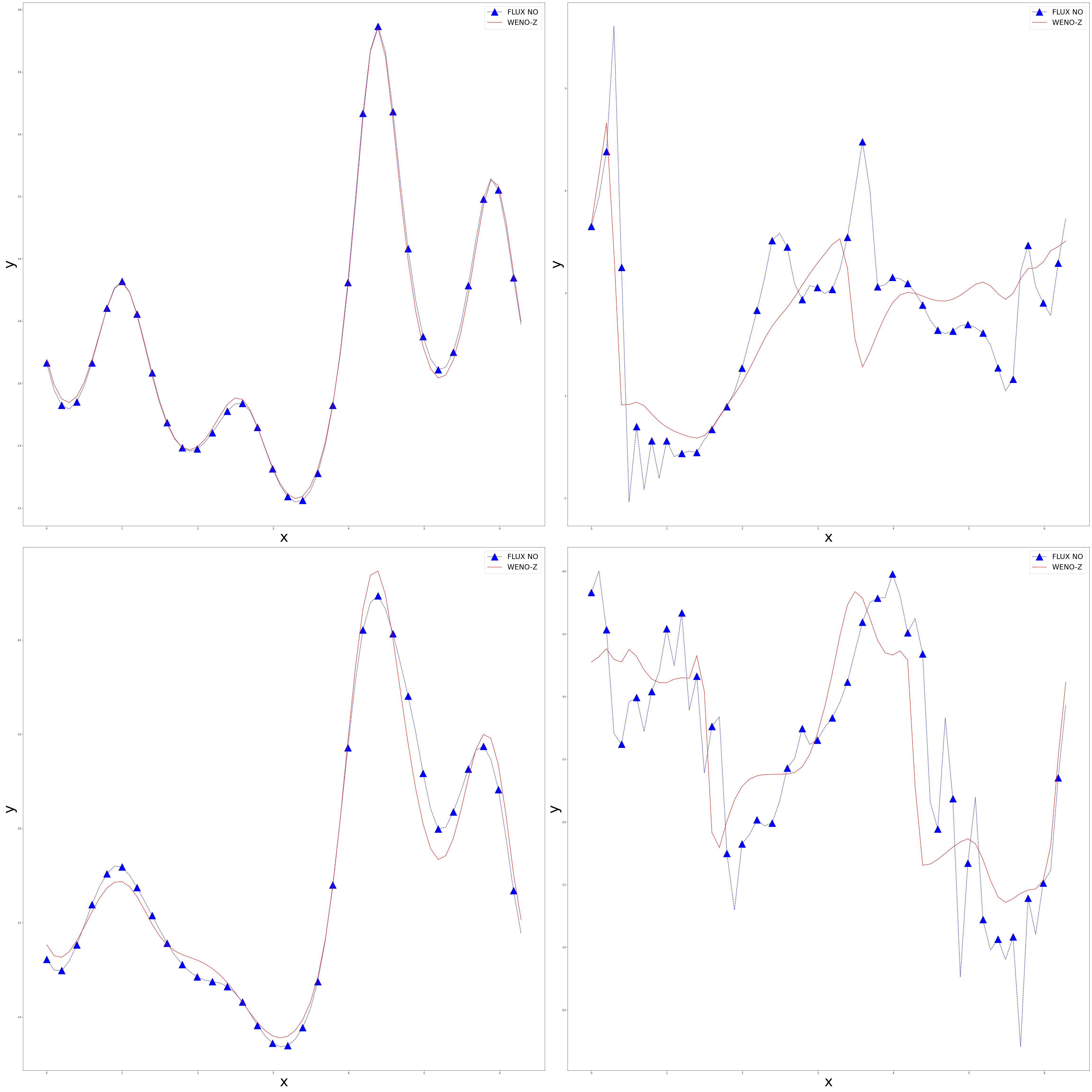}}%
  \caption{Time Evolution of Sectional Graphs of Flux NO and Reference Data for Each Component at Section $y=0.625\pi$. Snapshots taken at $t=0.5$ (top left), $t=0.75$ (bottom left), $t=2.0$ (top right), and $t=3.0$ (bottom right).}
  \label{2D_section_wzt}
\end{figure}

\begin{figure}
  \makebox[\textwidth][c]{\includegraphics[width=1.1\textwidth]{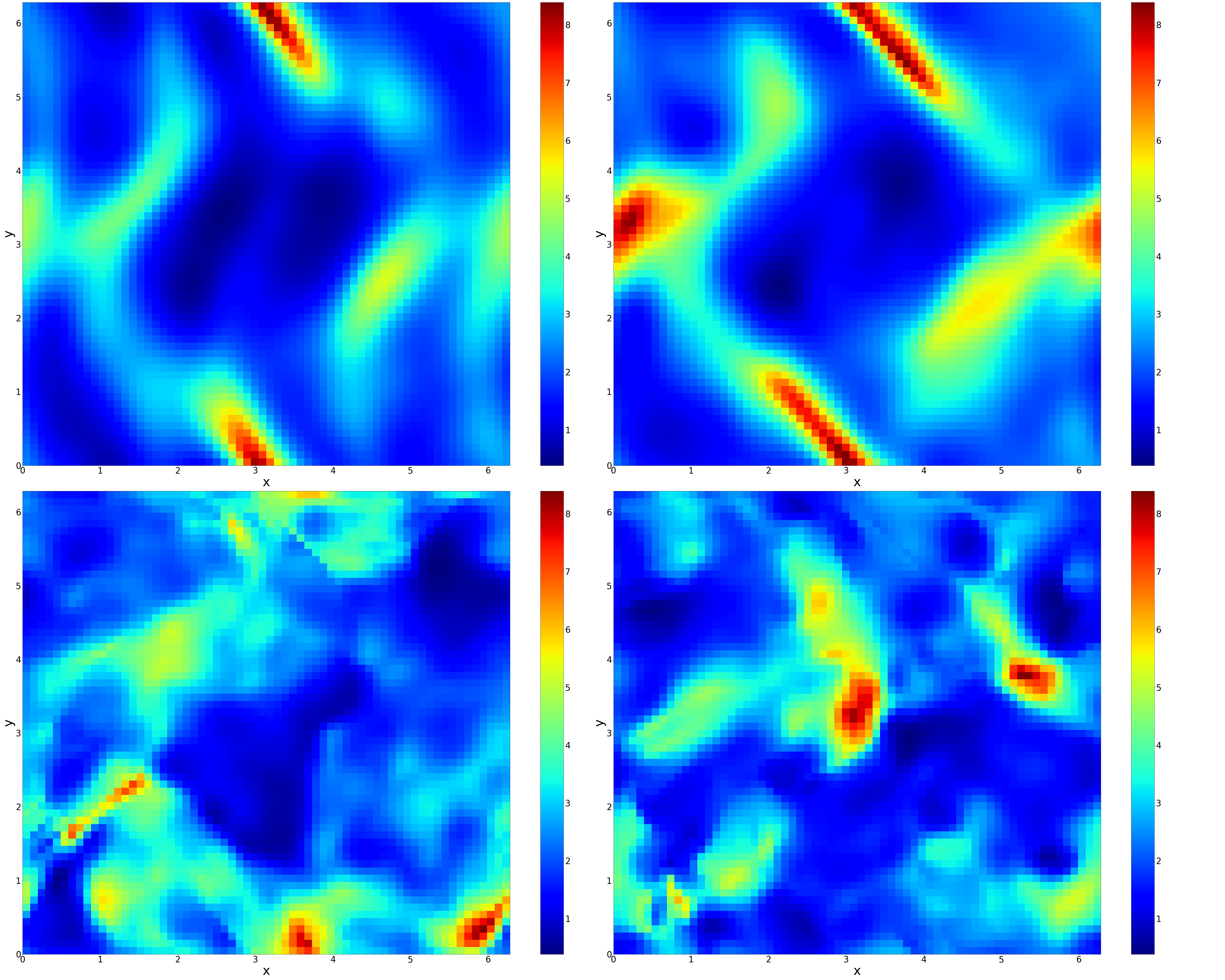}}%
  \caption{Snapshot of the Output from 3D FNO at $t=0.5$ (top left), $t=0.75$ (top right), $t=2.0$ (bottom left) and $t=3.0$ (bottom right) for the Orszag-Tang Vortex Problem.}
  \label{2D_short_wzt_3D}
\end{figure}

\section{Experimental Details}

All experiments were conducted using Pytorch 1.10.0, with Python 3.6.9. The specifications of the hardware environment are provided in Table \ref{tC}.

\begin{table}[H]
\small
\makebox[\textwidth]{
\begin{tabular}{|c|c|c|}
\hline
CPU & GPU & RAM \\
\hline
80 Intel(R) Xeon(R) Gold 6242R & 2 Nvidia RTX3090 & 503GB \\
\hline
\end{tabular}}
\caption{Specifications of Computer Hardware Used in the Study.} \label{tC}
\end{table}

\section{Courant--Friedrichs--Lewy Condition (CFL Condition)}

The Courant-Friedrichs-Lewy (CFL) condition is a crucial criterion for the stability of numerical solutions in computational simulations. It ensures that the physical propagation speed of a wave within a given timestep is less than or equal to the numerical propagation speed, which is essential for maintaining stability. For general N-dimensional case, the CFL condition is commonly expressed as follows:

\begin{equation*}
\Delta t \Bigg( \sum_{i=1}^{N} \frac{u_{i}}{\Delta x_{i}} \Bigg) \leq C
\end{equation*}
\noindent
Here, $u_{i}$ represents the wave speed, and $C$ is the Courant number, typically set to a value less than one. This constraint ensures that the numerical method can adequately capture the physical phenomena within the constraints of the time-step and spatial resolution.

\vskip 0.2in
\bibliography{FNO}
\nocite{Valiant:84,LeVeque:92,Pathak:22,shalev:88,Jin:95,Gopalani:22,Jaku:19,Kovachki:212,Wen:22,Abrahamsen:97,Alfven:42,Bar:21,Benitez:23,Bittencourt:04,Borges:08,Brackbill:80,Brio:88,Chen:22,Christlieb:14,Costa:07,Courant:52,Dai:98,Evans:88,Fu:19,Galtier:16,Godunov:59,Gottlieb:98,Gupta:21,Harten:87,Henrick:05,Hill:04,Holl:20,Holl:22,Jiang:96,Jiang:99,Kennel:85,Kim:24_2,Kovachki:21,Lax:54,Kim:24,Lee:23,Leer:74,Li:20,Li:21,Liu:94,Lu:21,Magiera:20,Mukhopadhyay:21,O'Shea:15,Pakmor:13,Pirozzoli:02,Priest:82,Ray:18,Rincon:19,Rosofsky:23,Rossmanith:06,Ruggeri:22,Shibata:11,Sweby:84,Toth:00,Vapnik:21,Vlasov:38,Wang:20,Wesson:78}

\end{document}